\documentclass[10pt,final,journal]{IEEEtran}
\usepackage[switch]{lineno} 
\usepackage{tabularx} 
\usepackage{booktabs}

\usepackage{stmaryrd}
\usepackage{amsthm} 
\usepackage{amsmath} 
\usepackage{bm}
\usepackage{graphicx}
\usepackage{epstopdf}
\usepackage{color}
\usepackage{float}
\usepackage{makecell}
\usepackage{mathtools}
\usepackage[colorlinks=false,linkcolor=black,urlcolor=black,bookmarksopen=true, hidelinks]{hyperref}

\usepackage{xcolor,soul}

\usepackage{cite}

\ifCLASSINFOpdf
\else
\fi
\usepackage{amsfonts,amssymb}

\usepackage{amsmath}
\usepackage{algorithm}
\usepackage{algpseudocode}
\ifCLASSOPTIONcompsoc
 \usepackage[caption=false,font=normalsize,labelfont=sf,textfont=sf]{subfig}
\else
 \usepackage[caption=false,font=footnotesize]{subfig}
\fi
\newtheorem{theorem}{Theorem}

\begin{document}
%

\title{A Closed-form Localization Method Utilizing Pseudorange Measurements from Two Non-synchronized Positioning Systems}

\author{Sihao Zhao, Xiao-Ping Zhang, Xiaowei Cui, and Mingquan Lu  
\thanks{This work was supported in part by the Natural Sciences and Engineering Research Council of Canada (NSERC), Grant No. RGPIN-2020-04661. \textit{(Corresponding author: Xiao-Ping Zhang.)}}
\thanks{S. Zhao, X.-P. Zhang are with the Department of Electrical, Computer and Biomedical Engineering, Ryerson University, Toronto, ON M5B 2K3, Canada (e-mail: sihao.zhao@ryerson.ca; xzhang@ryerson.ca).}
\thanks{X. Cui is with the Department of Electronic Engineering,
Tsinghua University, Beijing 100084, China (e-mail: cxw2005@tsinghua.edu.cn).}
\thanks{M. Lu is with the Department of Electronic Engineering,
Beijing National Research Center for Information Science and Technology, Tsinghua University, Beijing 100084, China. (e-mail: lumq@tsinghua.edu.cn).}
\thanks{Copyright (c) 20xx IEEE. Personal use of this material is permitted. However, permission to use this material for any other purposes must be obtained from the IEEE by sending a request to pubs-permissions@ieee.org.}}

\markboth{}%
{Shell \MakeLowercase{\textit{et al.}}: Bare Demo of IEEEtran.cls for IEEE Journals}
%




\maketitle
\begin{abstract}
In a time-of-arrival (TOA) or pseudorange based positioning system, user location is obtained by observing multiple anchor nodes (AN) at known positions. Utilizing more than one positioning systems, e.g., combining Global Positioning System (GPS) and BeiDou Navigation Satellite System (BDS), brings better positioning accuracy. However, ANs from two systems are usually synchronized to two different clock sources. Different from single-system localization, an extra user-to-system clock offset needs to be handled. Existing dual-system methods either have high computational complexity or sub-optimal positioning accuracy. In this paper, we propose a new closed-form dual-system localization (CDL) approach that has low complexity and optimal localization accuracy. We first convert the nonlinear problem into a linear one by squaring the distance equations and employing intermediate variables. Then, a weighted least squares (WLS) method is used to optimize the positioning accuracy. We prove that the positioning error of the new method reaches Cram\'er-Rao Lower Bound (CRLB) in far field conditions with small measurement noise. Simulations on 2D and 3D positioning scenes are conducted. Results show that, compared with the iterative approach, which has high complexity and requires a good initialization, the new CDL method does not require initialization and has lower computational complexity with comparable positioning accuracy. Numerical results verify the theoretical analysis on positioning accuracy, and show that the new CDL method has superior performance over the state-of-the-art closed-form method. Experiments using real GPS and BDS data verify the applicability of the new CDL method and the superiority of its performance in the real world.

\end{abstract} 

\begin{IEEEkeywords}
Time-of-arrival, pseudorange, Global Navigation Satellite System, closed-form localization, dual systems.
\end{IEEEkeywords}


%
\IEEEpeerreviewmaketitle

\section{Introduction}\label{Introduction}
%
%
%
%
\IEEEPARstart{P}{o}sition information is becoming more and more pivotal to many modern applications including smart cities, autonomous vehicles, Internet of Things (IoT), emergency rescues, \cite{ferreira2017localization,kuutti2018survey,liu2007survey, tahat2016look}. Among those pervasive positioning techniques, wireless localization systems are usually comprised of anchor nodes (AN) at known locations and user devices (UD) that need to be localized. Several measurement techniques including time-of-arrival (TOA), time-of-flight (TOF), angle-of-arrival (AOA), received signal strength (RSS), etc., can be adopted for localization \cite{guvenc2009survey,beuchat2019enabling,luo2017accuracy,yan2013review, shao2014efficient,feng2020kalman,tomic20163,wang2012novel,xue2019locate,hu2017robust,cai2019asynchronous}. The TOF requires perfect synchronization between the UD and the AN, which may be costly to obtain. The TOA or pseudorange measurement does not need such synchronization and is currently one of the most widely adopted methods to determine the UD position due to its relative device simplicity and localization accuracy. A typical example of such a scheme is the widely used Global Positioning System (GPS).

Positioning techniques based on pseudorange measurements from a single system are extensively studied in literature. They can be mainly categorized into two types, iterative methods and closed-form methods. Iterative methods based on Taylor series expansion are widely adopted \cite{kaplan2005understanding,foy1976position,teunissen1990nonlinear,borre2007software,zou2017iterative,zhao2016design}. They provide optimal positioning results that reach Cram\'er-Rao Lower Bound (CRLB). However, they require proper initialization and have high computational complexity.

A variety of closed-form localization methods, which have low computational complexity and require no initial guess, are developed. Schau and Robinson \cite{schau1987passive} employ the squared user distance as intermediate variable, and solve a quadratic equation to obtain the localization result. Smith and Abel  \cite{smith1987closed} use the same intermediate variable and equations and it only applies for over-determined cases. Chan and Ho  \cite{chan1994simple} propose a two-step weighted least squares (WLS) estimator that achieves CRLB at small noise level. Bancroft \cite{bancroft1985algebraic} employs the squared difference of the user position and clock offset as intermediate variable and obtains the localization result by finding the root of a quadratic equation containing this intermediate variable as unknown. Closed-form localization method based on the multidimensional scaling technique that utilizes a squared distance matrix are proposed in \cite{cheung2004least,so2007generalized}. However, all of the above methods and their improved versions such as \cite{cheung2006constrained,zhu2010joint,jiang2016multidimensional,chen2009novel,wei2009multidimensional,luo2019novel,song2019novel} are only applicable with measurements from synchronized ANs within a single system.

Utilizing more than one positioning systems provides navigation users with more measurements, and thus better availability and higher accuracy \cite{misra2006global,heng2014gnss,gao2012many ,zhao2014kalman}. For example, combining other global navigation satellite systems (GNSS) with GPS, such as Glonass, Galileo and BeiDou Navigation Satellite System (BDS), which are under development or becoming available, can provide better positioning services. However, these systems have different designs and thus have different clock bases \cite{langley2017introduction}. It causes positioning with multiple systems more challenging than in the single-system case. For the dual-system positioning case, iterative methods, which are modified from the single-system case by adding another clock offset term, are commonly adopted \cite{kaplan2005understanding,ma2019direct,gan2019combination}. However, they still require  proper initialization and have high complexity. A closed-form dual-system positioning algorithm is proposed by Juang and Tsai \cite{juang2009exact}, in which the positioning problem is converted to finding the solution of the two clock offset terms. However, this method does not provide optimal localization result (as will be shown later in this paper). Teng et al. \cite{teng2016closed} modify this method to simplify computation by reducing an unknown clock offset term. However, its estimate result is not optimal either. In addition, an extra measurement is required to reduce the clock term, making this method only applicable to over-determined cases, i.e., five instead of four measurements for 2D and six instead of five for 3D cases.

In this paper, we propose a new closed-form dual-system localization (CDL) method. We first difference the pseudorange measurements with a reference AN from the same system to remove the clock offset term and form the time-difference-of-arrival (TDOA) measurements. Squaring operation is taken on the distance equations, and two intermediate variables containing the distances between the unknown user position and the reference ANs are employed, to convert the non-linear problem into a linear one. After finding the solution of the two intermediate variables by solving a quadratic equation set, a WLS method is applied to obtain the user location. The covariance of the localization result is analyzed theoretically to evaluate its positioning accuracy. We prove that the analytic form of the localization error covariance is identical with CRLB under small measurement noise and far-field assumption. Simulations are conducted to compare the localization accuracy of the proposed new CDL method against existing representative methods. Numerical results show that the localization accuracy of the proposed CDL algorithm reaches CRLB under small noise and far field conditions, and is better than that of the state-of-the-art method in \cite{juang2009exact}. Furthermore, we conduct experiments using real GPS and BDS data. Results show the feasibility and performance of the new method in the real world. Compared with the iterative method, the new CDL method does not require initialization, and  the computational time reduces by about 40\% with similar positioning accuracy.

The paper is organized as follows. In Section II, the localization problem model for two non-synchronized systems is formulated. A new localization algorithm for the dual-system case named CDL is proposed in detail in Section III. Then the position error covariance is analyzed and compared with CRLB in Section IV. Simulations and real-data experiment are conducted to evaluate the performance of the new CDL method compared with other methods in Section V. Finally, Section VI draws the conclusion of this paper.

Main notations used in this paper are summarized in Table \ref{table_notation}.

\begin{table}[!t]
\caption{Notation List}
\label{table_notation}
\centering
\begin{tabular}{l p{5.5cm}}
\toprule
lowercase $x$&  scalar\\
bold lowercase $\boldsymbol{x}$ & vector\\
bold uppercase $\bm{X}$ & matrix\\
$\hat{x}$, $\hat{\boldsymbol{x}}$, $\hat{\bm{X}}$ & noisy version of a variable\\
$\tilde{x}$, $\tilde{\boldsymbol{x}}$, $\tilde{\bm{X}}$ & estimate of a variable\\
$\Vert \boldsymbol{x} \Vert$ & Euclidean norm of a vector\\
$\mathrm{tr}(\bm{X})$ & trace of a matrix\\
$[\bm{X}]_{i,:}$, $[\bm{X}]_{:,j}$ &the $i$-th row and the $j$-th column of a matrix, respectively\\
$[\bm{X}]_{i,j}$ &entry at the $i$-th row and the $j$-th column of a matrix\\
$[\boldsymbol{x}]_{i}$ &the $i$-th element of a vector\\
$\mathbb{E}[\cdot]$ & expectation operator \\
$\mathrm{diag}(\cdot)$ & diagonal matrix with the elements inside\\
$M$,$N$ & numbers of ANs of system $A$ and $B$, respectively\\
$\bm{O}_{M \times N}$ &$M \times N$ matrix with all-zero entries\\
$\bm{1}_M$ &  $M$-element vector filled with ones \\ $\bm{0}_M$ & $M$-element vector filled with zeros\\
$\boldsymbol{p}_{A_i}$, $\boldsymbol{p}_{B_j}$ & known position vectors of the $i$-th or $j$-th AN in system $A$ and $B$, respectively\\
$\boldsymbol{p}_u$ &  unknown position vector of the UD\\
$\rho_{A_i}$, $\rho_{B_j}$ & pseudorange measurements between the  UD and the $i$-th and $j$-th ANs in system $A$ and $B$, respectively\\
$r$ &  physical distance between the UD and AN\\
$b$ &  clock offset caused distance between the UD and AN\\
$\epsilon$ &  pseudorange measurement noise\\
$\sigma^2$ &  pseudorange measurement noise variance\\
$\boldsymbol{l}$ & unit line-of-sight (LOS) direction vector from the UD to AN\\
$\bm{F}$ &  Fisher information matrix\\
$\bm{W}$ & weighting matrix for CDL\\
$\bm{Q}$ & covariance matrix of TDOA measurements\\

\bottomrule
\end{tabular}
\end{table}

\section{Problem Statement}
\begin{figure}
	\centering
	\includegraphics[width=0.99\linewidth]{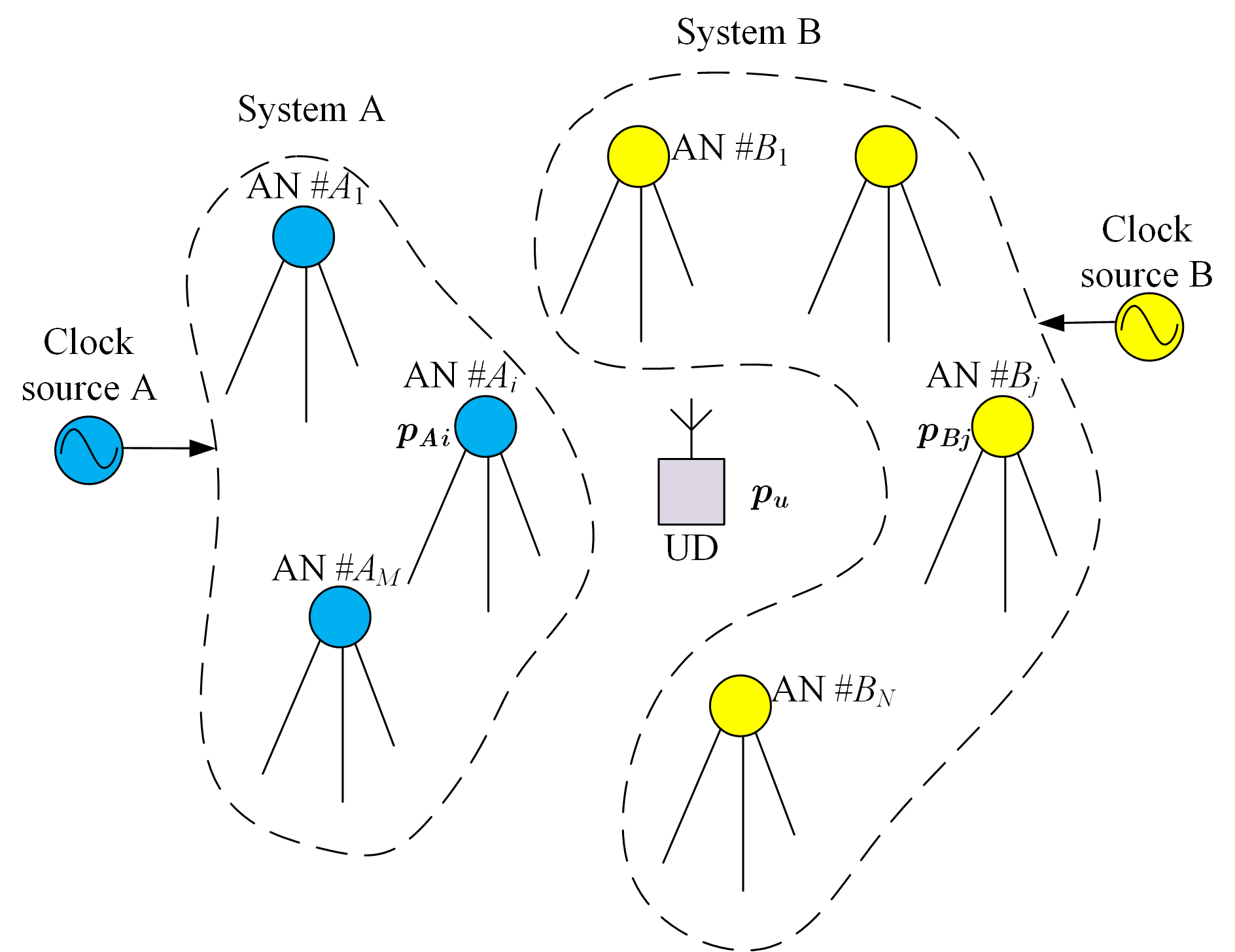} 
	\caption{Dual positioning systems setup. Systems $A$ and $B$ have independent clock sources. ANs have known positions. UD receives signals from ANs to localize itself.
	}
	\label{fig:systemsetup}
\end{figure}
We consider two positioning systems denoted as system $A$ and $B$, respectively, as depicted in Fig. \ref{fig:systemsetup}. System $A$ contains $M$ ANs and system $B$ contains $N$ ANs, i.e., $M$ and $N$ pseudorange measurements can be obtained from systems $A$ and $B$, respectively. The coordinates of all the ANs are known, which are denoted as $\boldsymbol{p}_{A_{i}}$ and $\boldsymbol{p}_{B_{j}}$, for the $i$-th AN in system $A$ and the $j$-th AN in system $B$, respectively, $i=1,\dotsc,M$ and $j=1,\dotsc,N$. The location of a UD, denoted as $\boldsymbol{p}_u$, is the unknown to be determined. The dimension of all the position vectors is $K$ (e.g., $K=2$ in 2D case and $K=3$ in 3D case), i.e., $\boldsymbol{p}_{A_{j}}, \boldsymbol{p}_{B_{j}}, \boldsymbol{p}_u\in \mathbb{R}^K$. Therefore, the distance between UD and AN are expressed by
\begin{equation} \label{eq:0202rvsposA}
r_{A_{i}} = \Vert\boldsymbol{p}_u-\boldsymbol{p}_{A_{i}}\Vert \text{,}
\end{equation}
and
\begin{equation} \label{eq:0202rvsposB}
r_{B_{j}} = \Vert\boldsymbol{p}_u-\boldsymbol{p}_{B_{j}}\Vert,
\end{equation}
where $r_{A_{i}}$ represents the true distance between UD and the $i$-th AN in system $A$, and $r_{B_{j}}$ represents the true distance between UD and the $j$-th AN in system $B$.

The ANs in both systems are synchronized to their own system clock source, i.e., clock sources $A$ and $B$ in Fig. \ref{fig:systemsetup}. However, the two systems are not synchronized, i.e., the two system clock sources are independent to each other.

For the system shown in Fig. \ref{fig:systemsetup}, there are usually two schemes to obtain TOA or pseudorange measurements. One is that the ANs transmit signals and the TOA is measured upon the UD reception. Another is a reverse to the first one, i.e., the UD transmits signal and ANs receive and measure the TOAs. Either way works and has real-world applications. Without loss of generality, in this paper we suppose that the system works based on the former scheme, which is adopted by GNSS, i.e., the ANs broadcast signals while a UD only receives. At the UD side, the TOA of the broadcast signal can be measured and thus a pseudorange measurement can be obtained by
\begin{equation} \label{eq:0201prA}
\rho_{A_{i}} =  r_{A_{i}} + b_A + \epsilon_{A_{i}},
\end{equation}
where $\rho_{A_{i}}$ is the pseudorange measurement between UD and the $i$-th AN in system $A$, $b_A$ is the range offset that equals to the product of the clock offset between UD and AN in system $A$ and the signal propagation speed, and $\epsilon_{A_{i}}$ is the i.i.d. pseudorange measurement noise that follows a zero mean Gaussian distribution with a variance of $\sigma_{A_{i}}^2$, i.e., $\epsilon_{A_{i}}\sim \mathcal{N}(0,\sigma_{A_{i}}^2)$. All the above variables have the same unit of meter.

Similarly, the pseudorange measurement in system $B$ is 
\begin{equation} \label{eq:0202prB}
\rho_{B_{j}} =  r_{B_{j}} + b_B + \epsilon_{B_{j}} \text{,}
\end{equation}
where the subscript ``$B$'' represents the corresponding variables in system $B$.


The aim of this dual-constellation localization problem is to find an accurate estimate of the user position $\boldsymbol{p}_u$ from the collection of the measurements with the relationship to the distances given by (\ref{eq:0202rvsposA}) to (\ref{eq:0202prB}). This is a non-linear problem. A new closed-form solution for this problem will be developed in the next section.

\section{A New Closed-form Dual-system Localization Method}
We develop a new closed-form Dual-system localization method (CDL) in this section. The proposed method has three key steps including linearization, identification of intermediate variables, and WLS localization for UD. The three steps are presented in detail in the following sub-sections.

\subsection{Linearization}
The unknowns of this dual-system localization problem contains two clock offset terms that are not of interest because we only need to determine the user position. We note that the clock bias is common within either individual system. Therefore, a natural idea is to remove them by differencing the pseudorange measurements with a common reference to form TDOA measurements.

Without loss of generality, the first AN in system $A$ and the first AN in system $B$ are selected as references. The differenced pseudorange measurement between the reference distance and other distances in system $A$ is given by 
\begin{equation} \label{eq:0301tdoaA}
\rho_{A_{i}}-\rho_{A_{1}} = r_{A_{i}A_{1}} + \epsilon_{A_{i}A_{1}},
\end{equation}
where
\begin{equation} \label{eq:0302rdoaA}
r_{A_{i}A_{1}}=r_{A_{i}}-r_{A_{1}} \text{,}
\end{equation}
and $\epsilon_{A_{i}A_{1}} = \epsilon_{A_{i}}-\epsilon_{A_{1}}$.


The relationship between the UD coordinates and the distance as given by (\ref{eq:0202rvsposA}) and (\ref{eq:0202rvsposB}) is nonlinear. In order to convert it to a linear relation, we take a square on $r_{A_{i}}$ in (\ref{eq:0302rdoaA}) and (\ref{eq:0202rvsposA}), and then come to
\begin{align} \label{eq:0302rAisquare}
r_{A_{i}}^2 &= r_{A_{i}A_{1}}^2+2r_{A_{i}A_1}r_{A_1} +r_{A_1}^2 \nonumber\\
&= \left\Vert \boldsymbol{p}_{A_i} \right\Vert ^2 - 2\boldsymbol{p}_{A_i}^T\boldsymbol{p}_u+\Vert \boldsymbol{p}_u\Vert ^2 \text{.}
\end{align}

In order to remove the squared term of the UD coordinates $\Vert \boldsymbol{p}_u\Vert ^2$, we substitute $i=1$ into (\ref{eq:0302rAisquare}), and obtain
\begin{equation} \label{eq:0303rA1square}
r_{A_1}^2 = \left\Vert \boldsymbol{p}_{A_1} \right\Vert ^2 - 2\boldsymbol{p}_{A_1}^T\boldsymbol{p}_u+\Vert \boldsymbol{p}_u\Vert ^2 \text{.}
\end{equation}

By subtracting (\ref{eq:0303rA1square}) from (\ref{eq:0302rAisquare}), the squared UD coordinates are removed and it reads
\begin{align} \label{eq:0304rAisquarevsPu}
&r_{A_iA_1}^2+2r_{A_iA_1}r_{A_1}  \nonumber\\
&= \left\Vert \boldsymbol{p}_{A_i} \right\Vert ^2 -\Vert \boldsymbol{p}_{A_1}\Vert ^2 + 2(\boldsymbol{p}_{A_1}^T-\boldsymbol{p}_{A_i}^T)\boldsymbol{p}_u \text{.}
\end{align}

We put the unknown UD position to the left and all the rest terms to the right, and (\ref{eq:0304rAisquarevsPu}) becomes
\begin{align} \label{eq:0305PuvsrA1}
&\left(\boldsymbol{p}_{A_1}^T-\boldsymbol{p}_{A_i}^T \right)\boldsymbol{p}_u\nonumber\\
&= r_{A_iA_1}r_{A_1} + \frac{1}{2}\left(r_{A_iA_1}^2 +  \left\Vert \boldsymbol{p}_{A_1} \right\Vert ^2 -\Vert \boldsymbol{p}_{A_i}\Vert ^2\right) \text{.}
\end{align}

Similarly, for system $B$, by replacing the subscript of ``$A$'' to ``$B$'', we have
\begin{align} \label{eq:0306PuvsrB1}
&\left(\boldsymbol{p}_{B_1}^T-\boldsymbol{p}_{B_j}^T \right)\boldsymbol{p}_u\nonumber\\
&= r_{B_jB_1}r_{B_1} + \frac{1}{2}\left(r_{B_jB_1}^2 +  \left\Vert \boldsymbol{p}_{B_1} \right\Vert ^2 -\Vert \boldsymbol{p}_{B_j}\Vert ^2\right) \text{.}
\end{align}

We use matrices and vectors to rewrite equations (\ref{eq:0305PuvsrA1}) and (\ref{eq:0306PuvsrB1}) into the collective form
\begin{equation} \label{eq:0307Puvsrvector}
\begin{split}
\bm{G}\boldsymbol{p}_u = \bm{C} \left[r_{A_1},  r_{B_1} \right]^T + \boldsymbol{h} \text{,}
\end{split}
\end{equation}
where
$$
\bm{G}=
\begin{bmatrix}
\boldsymbol{p}_{A_1}^T-\boldsymbol{p}_{A_2}^T \\
\vdots \\
\boldsymbol{p}_{A_1}^T-\boldsymbol{p}_{A_M}^T \\
\boldsymbol{p}_{B_1}^T-\boldsymbol{p}_{B_2}^T \\
\vdots \\
\boldsymbol{p}_{B_1}^T-\boldsymbol{p}_{B_N}^T \\
\end{bmatrix}
\text{,} \; \bm{C}=
\begin{bmatrix}
r_{A_2A_1} & 0 \\
\vdots & \vdots\\
r_{A_MA_1} & 0 \\
0 & r_{B_2B_1} \\
\vdots & \vdots\\
0 & r_{B_NB_1} \\
\end{bmatrix} \text{,}
$$
and
$$
\boldsymbol{h}= \frac{1}{2}
\begin{bmatrix}
r_{A_2A_1}^2+\left\Vert \boldsymbol{p}_{A_1} \right\Vert ^2 -\Vert \boldsymbol{p}_{A_2}\Vert ^2 \\
\vdots \\
r_{A_MA_1}^2+\left\Vert \boldsymbol{p}_{A_1} \right\Vert ^2 -\Vert \boldsymbol{p}_{A_M}\Vert ^2 \\
r_{B_2B_1}^2+\left\Vert \boldsymbol{p}_{B_1} \right\Vert ^2 -\Vert \boldsymbol{p}_{B_2}\Vert ^2 \\
\vdots \\
r_{B_NB_1}^2+\left\Vert \boldsymbol{p}_{B_1} \right\Vert ^2 -\Vert \boldsymbol{p}_{B_N}\Vert ^2 \\
\end{bmatrix}.
$$

At this stage, the linear relation of the unknown UD position $\boldsymbol{p}_{u}$ with the distance variables $r_{A_1}$ and $r_{B_1}$ is obtained in (\ref{eq:0307Puvsrvector}). The AN positions $\boldsymbol{p}_{A_{i}}$ and $\boldsymbol{p}_{B_{j}}$ are known and the differenced distances $r_{A_iA_1}$ and $r_{B_jB_1}$ can be approximated by $\rho_{A_{i}}-\rho_{A_{1}}$ and $\rho_{B_{j}}-\rho_{B_{1}}$, respectively. We treat these two distances of $r_{A_1}$ and $r_{B_1}$ as intermediate variables and find the solution of them, then the UD position can be computed using this linear relation of (\ref{eq:0307Puvsrvector}).

\subsection{Identification of Intermediate Variables}
The intermediate variables $r_{A_1}$ and $r_{B_1}$ will be solved in this sub-section. First, by observing (\ref{eq:0303rA1square}), we note that if $\boldsymbol{p}_u$ is replaced by $r_{A_1}$ and $r_{B_1}$, an equation set with respect to the intermediate variables can be formed and solved. To this end, we then express the UD position by
\begin{equation} \label{eq:0317LSnoweight}
\begin{split}
\boldsymbol{p}_u = \left(\bm{G}^T\bm{G}\right)^{-1}\bm{G}^T\left(\bm{C} \left[r_{A_1},  r_{B_1} \right]^T + \boldsymbol{h}\right) \text{,}
\end{split}
\end{equation}
where $\bm{G}$ has full column rank which is usually satisfied when there are sufficient amount of ANs with a proper geometry.



By substituting $\boldsymbol{p}_{u}$ from (\ref{eq:0317LSnoweight}) into (\ref{eq:0303rA1square}), a quadratic equation with the two intermediate variables is formed. We replace the subscript of $A_1$ in (\ref{eq:0303rA1square}) with $B_1$, and substitute $\boldsymbol{p}_{u}$ from (\ref{eq:0317LSnoweight}) into it again, another quadratic equation with the same two variables is obtained. These two quadratic equations are given by
\begin{equation} \label{eq:0319quadratic1}
\begin{split}
a_1 r_{A_1}^2+b_1 r_{A_1}r_{B_1} +c_1 r_{B_1}^2 + d_1 r_{A_1}+e_1 r_{B_1}+f_1=0 \text{,}
\end{split}
\end{equation}
and
\begin{equation} \label{eq:0320quadratic2}
\begin{split}
a_2 r_{A_1}^2+b_2 r_{A_1}r_{B_1} +c_2 r_{B_1}^2 + d_2 r_{A_1}+e_2 r_{B_1}+f_2=0 \text{,}
\end{split}
\end{equation}
where
$$
a_1=[\bm{S}]_{:,1}^T[\bm{S}]_{:,1}-1, \; b_1=b_2=2[\bm{S}]_{:,1}^T[\bm{S}]_{:,2},
$$
$$
c_1=[\bm{S}]_{:,2}^T[\bm{S}]_{:,2}, \;
d_1=2[\bm{S}]_{:,1}(\boldsymbol{g}-\boldsymbol{p}_{A_1}),
$$
$$
e_1=2[\bm{S}]_{:,2}(\boldsymbol{g}-\boldsymbol{p}_{A_1}),\;
f_1=\left(\boldsymbol{g}-\boldsymbol{p}_{A_1}\right)^T\left(\boldsymbol{g}-\boldsymbol{p}_{A_1}\right),
$$
$$
a_2=[\bm{S}]_{:,1}^T[\bm{S}]_{:,1},\;
c_2=[\bm{S}]_{:,2}^T[\bm{S}]_{:,2}-1,
$$
$$
d_2=2[\bm{S}]_{:,1}(\boldsymbol{g}-\boldsymbol{p}_{B_1}),\;
e_2=2[\bm{S}]_{:,2}(\boldsymbol{g}-\boldsymbol{p}_{B_1}),
$$
$$
f_2=\left(\boldsymbol{g}-\boldsymbol{p}_{B_1}\right)^T\left(\boldsymbol{g}-\boldsymbol{p}_{B_1}\right),
$$
in which the matrix $\bm{S}$ and vector $\boldsymbol{g}$ are defined as
$$
\bm{S} \triangleq \left(\bm{G}^T\bm{G}\right)^{-1}\bm{G}^T\bm{C} \text{,}
$$
and
$$
\boldsymbol{g} \triangleq \left(\bm{G}^T\bm{G}\right)^{-1}\bm{G}^T\boldsymbol{h}\text{.}
$$

The quadratic equation set of (\ref{eq:0319quadratic1}) and (\ref{eq:0320quadratic2}) can be solved analytically. The approach is given in Appendix \ref{Appendix1}. There are at most 4 sets of roots. We know that the intermediate variables $r_{A_1}$ and $r_{B_1}$ represent the distances between the UD and the ANs. They are thereby real and non-negative values. Select these real and non-negative roots as reasonable solutions to $r_{A_1}$ and $r_{B_1}$.

\subsection{WLS Localization}

After obtaining the intermediate variables $r_{A_1}$ and $r_{B_1}$, we can estimate $\boldsymbol{p}_{u}$ in the expression of the intermediate variables by applying a WLS method to (\ref{eq:0307Puvsrvector}), and it comes to
\begin{equation} \label{eq:0309PuWLS}
\begin{split}
\tilde{\boldsymbol{p}}_u = \left(\bm{G}^T\bm{W}^{-1}\bm{G}\right)^{-1}\bm{G}^T\bm{W}^{-1}\left(\bm{C} \left[\tilde{r}_{A_1},  \tilde{r}_{B_1} \right]^T + \boldsymbol{h}\right) \text{,}
\end{split}
\end{equation}
where $\tilde{r}_{A_1}$ and $\tilde{r}_{B_1}$ represent the solutions from (\ref{eq:0319quadratic1}) and (\ref{eq:0320quadratic2}), $\tilde{\boldsymbol{p}}_u$ represents the position result estimated from $\tilde{r}_{A_1}$ and $\tilde{r}_{B_1}$, and $\bm{W}$ is the weighting matrix.

\begin{theorem}
	Under the condition of far field and small measurement noise, i.e., the squared error term $o(\epsilon^2)$ is negligible, the weighting matrix $\bm{W}$ in (\ref{eq:0309PuWLS}) has the form of 
	\begin{equation} \label{eq:0316Weightmatrix}
	\bm{W}=\bm{D}\bm{Q}\bm{D} \text{,}
	\end{equation}
	where
	$
	\bm{D}= \mathrm{diag} \left(r_{A_2},\cdots,r_{A_M},r_{B_2},\cdots,r_{B_N}\right)
	$,
	

	$$
	\bm{Q}=
	\begin{bmatrix}
	\bm{Q}_A
	& \bm{O}_{(M-1) \times (N-1)} \\
	\bm{O}_{(N-1) \times (M-1)} &
	\bm{Q}_B
	\end{bmatrix}\text{,}
	$$
	$$
	\bm{Q}_A =
	\left[
	\begin{matrix}
	\sigma_{A_1}^2+ \sigma_{A_2}^2 &\sigma_{A_1}^2 & \cdots &\sigma_{A_1}^2  \\
	\sigma_{A_1}^2 &  \sigma_{A_1}^2+ \sigma_{A_3}^2 &\cdots &\sigma_{A_1}^2\\
	\vdots & \vdots & \ddots & \vdots \\
	\sigma_{A_1}^2 & \sigma_{A_1}^2 & \cdots &\sigma_{A_1}^2+ \sigma_{A_M}^2
	\end{matrix}
	\right]\text{,}
	$$
	and
	$$
	\bm{Q}_B =
	\left[
	\begin{matrix}
	\sigma_{B_1}^2+ \sigma_{B_2}^2 &\sigma_{B_1}^2 & \cdots &\sigma_{B_1}^2  \\
	\sigma_{B_1}^2 & \sigma_{B_1}^2+ \sigma_{B_3}^2 & \cdots & \sigma_{B_1}^2 \\
	\vdots & \vdots & \ddots & \vdots \\
	\sigma_{B_1}^2 & \sigma_{B_1}^2 &\cdots &\sigma_{B_1}^2+ \sigma_{B_N}^2
	\end{matrix}
	\right] \text{.}
	$$	
\end{theorem}
\textit{Proof.} See Appendix \ref{Appendix2}.

Theorem 1 gives the construction method for the weighting matrix $\bm{W}$ in (\ref{eq:0309PuWLS}). We note that this weighting matrix is related to the distances between UD and ANs, which form the matrix $\bm{D}$ in (\ref{eq:0316Weightmatrix}). A natural way to compute these distances in matrix $\bm{D}$ is using the UD position. However, at this stage, the UD position has not been found yet. Instead, the intermediate variables representing the distances from the UD to the reference ANs have been solved from the previous step. Under the condition of far-field and small measurement noise, the measurement noise term in (\ref{eq:0301tdoaA}) is at least one order of magnitude smaller than the distances between UD and ANs. Therefore, the distance-related entries in $\bm{D}$ can be approximated by the roots of the intermediate variables and the pseudorange measurements as

\begin{equation} \label{eq:0320estimatedistanceA}
r_{A_i} \approx \tilde{r}_{A_1} + \rho_{A_i}-\rho_{A_1}, \; i=2,\cdots,M\text{,}
\end{equation}
and
\begin{equation} \label{eq:0321estimatedistanceB}
r_{B_j} \approx \tilde{r}_{B_1} + \rho_{B_j}-\rho_{B_1}, \; j=2,\cdots,N.
\end{equation}

With the above estimated distances, the weighting matrix $\bm{W}$ can be computed using (\ref{eq:0316Weightmatrix}). Then, we apply (\ref{eq:0309PuWLS}) to obtain the position estimate. 

It is possible that there are multiple position solutions due to multiple reasonable roots for $r_{A_1}$ and $r_{B_1}$. Furthermore, we use the weighted sum of the squared residual of TDOA as a selection criterion for the final solution, i.e., the position estimate that minimizes this sum is selected as the final result. The selection strategy is given by

\begin{equation} \label{eq:0322residual}
\begin{split}
\min_{\tilde{\boldsymbol{p}}_u} \boldsymbol{d}_{\rho}^T\bm{Q}^{-1}\boldsymbol{d}_{\rho} \text{,}
\end{split}
\end{equation}
where $\boldsymbol{d}_{\rho}$ is a vector containing all the residuals as given by
$$
\left[\boldsymbol{d}_{\rho}\right]_i=\left\{
\begin{array}{ll}
\rho_{A_{i+1}}-\rho_{A_1}-\Vert\tilde{\boldsymbol{p}}_u-\boldsymbol{p}_{A_{i+1}}\Vert +\Vert\tilde{\boldsymbol{p}}_u-\boldsymbol{p}_{A_1}\Vert,
\\
\hspace{4.7cm} i=1,\cdots,M-1\\
\rho_{B_{i-M+2}}-\rho_{B_1}-\Vert\tilde{\boldsymbol{p}}_u-\boldsymbol{p}_{B_{i-M+2}}\Vert +\Vert\tilde{\boldsymbol{p}}_u-\boldsymbol{p}_{B_1}\Vert
,\\
\hspace{3.8cm}i=M,\cdots,M+N-2 \text{.}
\end{array}
\right. \text{.}
$$

When the number of ANs is large and all measurement noise variances are identical, (\ref{eq:0322residual}) can reduce to a simplified form of $\min_{\tilde{\boldsymbol{p}}_u} \boldsymbol{d}_{\rho}^T\boldsymbol{d}_{\rho}$, which saves computation. The derivation of this simplified selection strategy is given in Appendix \ref{Appendix3}.



The entire procedure of the proposed new method is summarized in Algorithm 1.

\begin{algorithm} 
	\caption{Closed-form Dual-system Localization (CDL)}
	\begin{algorithmic}[1]
		\State Input pseudorange measurements $\rho_{A_i}$ and $\rho_{B_j}$,  and AN positions $\boldsymbol{p}_{A_i}$, $ i=1,\cdots,M$, and $\boldsymbol{p}_{B_j}$,  $j=1,\cdots,N$.
		\State Linearization:
		Form matrix $\bm{G}$ and $\bm{C}$ and vector $\boldsymbol{h}$ based on (\ref{eq:0301tdoaA}) and (\ref{eq:0307Puvsrvector}).
		\State Identification of intermediate variables: Solve quadratic equations (\ref{eq:0319quadratic1}) and (\ref{eq:0320quadratic2}) and select the real and non-negative root(s).
		\State WLS localization: Compute candidate position results using (\ref{eq:0309PuWLS}). Select position result $\tilde{\boldsymbol{p}}_u$ that minimizes (\ref{eq:0322residual}).
		\State Output the selected position result.
	\end{algorithmic}
\end{algorithm}

\section{Error Analysis}\label{PositionError}
The Cram\'er-Rao Lower Bound (CRLB) is usually used to evaluate the error variance of an unbiased estimator. In this section, we derive the CRLB of the dual-system localization case and compare the covariance of the localization error from the proposed new CDL method against CRLB.
\subsection{CRLB for Dual-system Localization}
The CRLB of the dual-system localization problem using TDOA measurements is derived as a benchmark. The CRLB relating to the parameter vector $\boldsymbol{\theta}$  is defined as
\begin{equation} \label{eq:CRLB_Fisher}
\mathsf{CRLB}(\boldsymbol{\theta})\triangleq\bm{F}^{-1}(\boldsymbol{\theta}) \text{,}
\end{equation}
where $\bm{F}$ is the Fisher information matrix (FIM),  and in the dual-system localization case, the parameter to be estimated is the user position, i.e., $\boldsymbol{\theta}=\boldsymbol{p}_u$.

The entry of FIM is expressed by
\begin{equation} \label{eq:FisherExpectation}
[\bm{F}(\boldsymbol{\theta})]_{u,v}=-\mathbb{E}\left[\frac{\partial^2 \ln p(\boldsymbol{\rho}_D|\boldsymbol{\theta})}{\partial[\boldsymbol{\theta}]_u \partial[\boldsymbol{\theta}]_v}\right] \text{,}
\end{equation}
in which $p(\boldsymbol{\rho}_D|\boldsymbol{\theta})$ is the likelihood function, and $\boldsymbol{\rho}_D$ is a vector containing all TDOA measurements.

Therefore, when using TDOA measurements from dual systems as given by (\ref{eq:0301tdoaA}), the likelihood function is written as
\begin{equation} \label{eq:likelihood}
p(\boldsymbol{\rho}_D|\boldsymbol{\theta}) = \frac{\exp\left(-\frac{1}{2}\boldsymbol{f}(\boldsymbol{\theta})^T\bm{Q}^{-1}\boldsymbol{f}(\boldsymbol{\theta})\right)}{\sqrt{(2\pi)^{M+N}|\bm{Q}|}} \text{.}
\end{equation}
where
$$[\boldsymbol{f}(\boldsymbol{\theta})]_i=
\left\{
\begin{array}{ll}
\rho_{A_{i+1}}-\rho_{A_1}-\Vert{\boldsymbol\theta}-\boldsymbol{p}_{A_{i+1}}\Vert +\Vert{\boldsymbol{\theta}}-\boldsymbol{p}_{A_1}\Vert, \\ \hspace{4.3cm}i=1,\cdots,M-1 \\
\rho_{B_{i-M+2}}-\rho_{B_1}-\Vert{\boldsymbol\theta}-\boldsymbol{p}_{B_{i-M+2}}\Vert +\Vert{\boldsymbol{\theta}}-\boldsymbol{p}_{A_1}\Vert, \\ \hspace{3.4cm}i=M,\cdots,M+N-2 \text{.}
\end{array}
\right. \text{.}
$$

Then we have
\begin{equation} \label{eq:0403Expectation}
-\mathbb{E}\left[\frac{\partial^2 \ln p(\boldsymbol{\rho}_D|\boldsymbol{\theta})}{\partial\boldsymbol{\theta} \partial\boldsymbol{\theta}^T}\right]=
\left(\frac{\partial{\boldsymbol{f}(\boldsymbol{\theta})}}{\partial{\boldsymbol{\theta}}}\right)^T\bm{Q}^{-1}\frac{\partial\boldsymbol{f}(\boldsymbol{\theta})}{\partial{\boldsymbol{\theta}}} \text{.}
\end{equation}


The row of the first-order derivative of function $\boldsymbol{f}(\boldsymbol{\theta})$ is written as
\begin{equation} \label{eq:0405Hmatrix}
\left[\frac{\partial\boldsymbol{f}(\boldsymbol{\theta})}{\partial \boldsymbol{\theta}}\right]_{i,:} =
\left\{
\begin{matrix*}[l]
\boldsymbol{l}_{A_1}^T-\boldsymbol{l}_{A_{i+1}}^T, & \hspace{0.9cm} i=1,\cdots,M-1 \\
\boldsymbol{l}_{B_1}^T-\boldsymbol{l}_{B_{i-M+2}}^T,&i=M,\cdots,M+N-2\text{,}
\end{matrix*} 
\right.
\end{equation}
where $\boldsymbol{l}$ is the unit line-of-sight (LOS) direction vector from the UD to AN, and
$$\boldsymbol{l}_{A_i}=\frac{\boldsymbol{p}_{A_i} - {\boldsymbol{p}}_u}{\Vert \boldsymbol{p}_{A_i} - {\boldsymbol{p}}_u\Vert } \text{,}$$ and  $$\boldsymbol{l}_{B_i}=\frac{\boldsymbol{p}_{B_i} - {\boldsymbol{p}}_u}{\Vert \boldsymbol{p}_{B_i} - {\boldsymbol{p}}_u\Vert } \text{.}$$

At this stage, the CRLB using TDOA measurements from dual systems is obtained in (\ref{eq:0403Expectation}). When directly using TOA or pseudorange measurements, the position related CRLB is identical with the CRLB using TDOA. It is proved in Appendix \ref{Appendix4}.

\subsection{Localization Error Covariance of CDL}
We denote the localization error vector by $\Delta\boldsymbol{p}_u$, and the distance errors by $\Delta r_{A_1}$ and $\Delta r_{B_1}$, respectively. When there are measurement noises, (\ref{eq:0307Puvsrvector}) becomes
\begin{equation} \label{eq:0408disturbPu}
\begin{split}
\bm{G}\left(\boldsymbol{p}_u +\Delta\boldsymbol{p}_u\right)= \hat{\bm{C}} \left[r_{A_1}+\Delta r_{A_1},  r_{B_1}+\Delta r_{B_1} \right]^T + \hat{\boldsymbol{h}} \text{,}
\end{split}
\end{equation}
where $\hat{\bm{C}}$ and $\hat{\boldsymbol{h}}$ are the noisy version of  ${\bm{C}}$ and ${\boldsymbol{h}}$, respectively, as given by

$$
\hat{\bm{C}}=
\begin{bmatrix}
r_{A_2A_1}+\epsilon_{A_2A_1} & 0 \\
\vdots & \vdots\\
r_{A_MA_1}+\epsilon_{A_MA_1} & 0 \\
0 & r_{B_2B_1}+\epsilon_{B_2B_1} \\
\vdots & \vdots\\
0 & r_{B_NB_1}+\epsilon_{B_NB_1} \\
\end{bmatrix} 
\text{,}
$$
and
$$
\hat{\boldsymbol{h}}= \frac{1}{2}
\begin{bmatrix}
(r_{A_2A_1}+\epsilon_{A_2A_1})^2+\left\Vert \boldsymbol{p}_{A_1} \right\Vert ^2 -\Vert \boldsymbol{p}_{A_2}\Vert ^2 \\
\vdots \\
(r_{A_MA_1}+\epsilon_{A_MA_1})^2+\left\Vert \boldsymbol{p}_{A_1} \right\Vert ^2 -\Vert \boldsymbol{p}_{A_M}\Vert ^2 \\
(r_{B_2B_1}+\epsilon_{B_2B_1})^2+\left\Vert \boldsymbol{p}_{B_1} \right\Vert ^2 -\Vert \boldsymbol{p}_{B_2}\Vert ^2 \\
\vdots \\
(r_{B_NB_1}+\epsilon_{B_NB_1})^2+\left\Vert \boldsymbol{p}_{B_1} \right\Vert ^2 -\Vert \boldsymbol{p}_{B_N}\Vert ^2 \\
\end{bmatrix}.
$$

Without loss of generality, the first element of (\ref{eq:0408disturbPu}) can be derived as
\begin{align} \label{eq:0409disturbPurow1}
&\left[\bm{G}\left(\boldsymbol{p}_u +\Delta\boldsymbol{p}_u\right)\right]_{1} \nonumber \\
&=[\hat{\bm{C}}]_{1,:}\left[r_{A_1}+\Delta r_{A_1},  r_{B_1}+\Delta r_{B_1} \right]^T+[\hat{\boldsymbol{h}}]_1 \nonumber \\
&= [r_{A_2A_1}+\epsilon_{A_2A_1}, 0] \left[r_{A_1}+\Delta r_{A_1},  r_{B_1}+\Delta r_{B_1} \right]^T \nonumber \\
& \;\;\;\; + \frac{1}{2}(r_{A_2A_1}+\epsilon_{A_2A_1})^2+\frac{1}{2}\left(\Vert \boldsymbol{p}_{A_1}\Vert ^2-\Vert \boldsymbol{p}_{A_2}\Vert ^2 \right)\nonumber \\
&= r_{A_1}r_{A_2A_1}+r_{A_2A_1}\Delta r_{A_1}+r_{A_1}\epsilon_{A_2A_1}+\epsilon_{A_2A_1}\Delta r_{A_1} \nonumber \\
&\;\;\;\;  +\frac{1}{2}(r_{A_2A_1}^2+\Vert\boldsymbol{p}_{A_1}\Vert ^2-\Vert \boldsymbol{p}_{A_2}\Vert ^2)\nonumber \\
& \;\;\;\;+r_{A_2A_1}\epsilon_{A_2A_1}+\frac{1}{2}\epsilon_{A_2A_1}^2 \text{.}
\end{align}

By subtracting
$$
\left[\bm{G}\boldsymbol{p}_u \right]_{1}=r_{A_1}r_{A_2A_1}+\frac{1}{2}\left(r_{A_2A_1}^2+\Vert\boldsymbol{p}_{A_1}\Vert ^2-\Vert \boldsymbol{p}_{A_2}\Vert ^2\right) \text{,}
$$
which is the first element of (\ref{eq:0307Puvsrvector}), from (\ref{eq:0409disturbPurow1}), we come to
\begin{align} \label{eq:0411deltaPu}
&\left[\bm{G}\Delta\boldsymbol{p}_u\right]_{1}\nonumber\\
&= r_{A_2A_1}\Delta r_{A_1}+r_{A_2}\epsilon_{A_2A_1}+\epsilon_{A_2A_1}\Delta r_{A_1} +\frac{1}{2}\epsilon_{A_2A_1}^2 \text{.}
\end{align}

Given the condition of small measurement noise and far field, the distance error $\Delta r_{A_1}$ is equal to the projection of the UD position error $\Delta\boldsymbol{p}_u$ onto the line-of-sight (LOS) direction. This relationship is given by
\begin{equation} \label{eq:0414drvsdPu}
\Delta r_{A_1}=-\boldsymbol{l}_{A_1}^T \Delta\boldsymbol{p}_u \text{.}
\end{equation}

We substitute (\ref{eq:0414drvsdPu}) into (\ref{eq:0411deltaPu}), expand $\left[\bm{G}\Delta\boldsymbol{p}_u\right]_{1}$, ignore the quadratic error terms, move all the $\Delta\boldsymbol{p}_u$ terms to the left of the equation, and then come to
\begin{align} \label{eq:0416deltaPu1}
&\left[\bm{G}\Delta\boldsymbol{p}_u\right]_{1}+r_{A_2A_1}\boldsymbol{l}_{A_1}^T \Delta\boldsymbol{p}_u \nonumber\\
&=\left(\boldsymbol{p}_{A_1}^T-\boldsymbol{p}_{A_2}^T\right)\Delta\boldsymbol{p}_u+(r_{A_2}-r_{A_1})\boldsymbol{l}_{A_1}^T \Delta\boldsymbol{p}_u \nonumber\\
&=\left(\boldsymbol{p}_{A_1}^T-\boldsymbol{p}_{u}^T-\left(\boldsymbol{p}_{A_2}^T-\boldsymbol{p}_{u}^T\right)\right)\Delta\boldsymbol{p}_u+(r_{A_2}-r_{A_1})\boldsymbol{l}_{A_1}^T \Delta\boldsymbol{p}_u \nonumber\\
&=(r_{A_1}\boldsymbol{l}_{A_1}^T-r_{A_2}\boldsymbol{l}_{A_2}^T)\Delta\boldsymbol{p}_u+(r_{A_2}-r_{A_1})\boldsymbol{l}_{A_1}^T \Delta\boldsymbol{p}_u \nonumber\\
&= r_{A_2} (\boldsymbol{l}_{A_1}^T-\boldsymbol{l}_{A_2}^T)\Delta\boldsymbol{p}_u \nonumber\\
&=r_{A_2}\epsilon_{A_2A_1} \text{.}
\end{align}

The distance error $\Delta r_{B_1}$ is treated similarly as (\ref{eq:0414drvsdPu}), and thus the other elements of $\bm{G}\Delta\boldsymbol{p}_u$ can be processed similarly as (\ref{eq:0416deltaPu1}). After eliminating the $r_{A_2}$ term on both sides of (\ref{eq:0416deltaPu1}), we then write it in vector form as
\begin{equation} \label{eq:0417delPuvector}
\bm{H}\Delta\boldsymbol{p}_u=\boldsymbol{\epsilon}_D \text{,}
\end{equation}
where
$$
\bm{H}=
\begin{bmatrix}
\boldsymbol{l}_{A_1}^T-\boldsymbol{l}_{A_2}^T \\
\vdots \\
\boldsymbol{l}_{A_1}^T-\boldsymbol{l}_{A_M}^T \\
\boldsymbol{l}_{B_1}^T-\boldsymbol{l}_{B_2}^T \\
\vdots \\
\boldsymbol{l}_{B_1}^T-\boldsymbol{l}_{B_N}^T \\
\end{bmatrix}
, \; 
\boldsymbol{\epsilon}_D = 
\begin{bmatrix}
\epsilon_{A_2A_1} \\
\vdots \\
\epsilon_{A_MA_1} \\
\epsilon_{B_2B_1} \\
\vdots \\
\epsilon_{B_NB_1} \\
\end{bmatrix} \text{.}
$$

We notice that the covariance of $\boldsymbol{\epsilon}_D$ is given by $\bm{Q}$ as defined in (\ref{eq:0316Weightmatrix}). Hence, the covariance of $\Delta\boldsymbol{p}_u$ is written as
\begin{equation} \label{eq:0419covdelPu}
\mathbb{E}[\Delta\boldsymbol{p}_u\Delta\boldsymbol{p}_u^T]=\left(\bm{H}^T\bm{Q}^{-1}\bm{H}\right)^{-1} \text{.}
\end{equation}

It can be observed that (\ref{eq:0419covdelPu}) is identical with the inverse of (\ref{eq:0403Expectation}). Thus, we have proved that the proposed new CDL method reaches CRLB under the condition of far field and small measurement noise.

\section{Performance Evaluation}
In this section, after the evaluation metrics are briefly introduced, simulation tests as well as real data experiment are carried out to evaluate the performance of the new CDL method. The iterative method using the TOAs \cite{borre2007software}, which is commonly adopted in many applications such as GNSS receivers, is selected as one of the comparison methods. The state-of-the-art closed-form dual-system method proposed by Juang and Tsai\cite{juang2009exact} (referred to as Juang's method hereinafter) is implemented as another comparison. The computational platform running the following simulations is Matlab R2019b on a PC with Intel Core i5-4590 CPU @3.3GHz and 32GB RAM.

\subsection{Localization Performance Metrics}
The root mean square error (RMSE) of the localization results is used to evaluate the positioning accuracy in the simulation tests. It is given by
\begin{equation} \label{eq:RMSEdef}
RMSE=\sqrt{\frac{1}{N_s}\sum_{1}^{N_s}\Vert\boldsymbol{p}_u-\tilde{\boldsymbol{p}}_u\Vert^2} \text{,}
\end{equation}
where $N_s$ is the total number of simulation runs.

CRLB is used as a benchmark to assess the localization accuracy. In this 2D scene, the position error lower bound derived from CRLB is written as
\begin{equation} \label{eq:0505CRLB2D}
\begin{split}
error_{LB}=\sqrt{\mathsf{CRLB}\left(\left[\boldsymbol{p}_u\right]_1\right)+\mathsf{CRLB}\left([\boldsymbol{p}_u]_2\right)} \text{.}
\end{split}
\end{equation}

For 3D cases, the position error bound is similar to (\ref{eq:0505CRLB2D}) but the term representing the third axis is added.

\subsection{2D Simulation}
\begin{figure}
	\centering
	\includegraphics[width=0.9\linewidth]{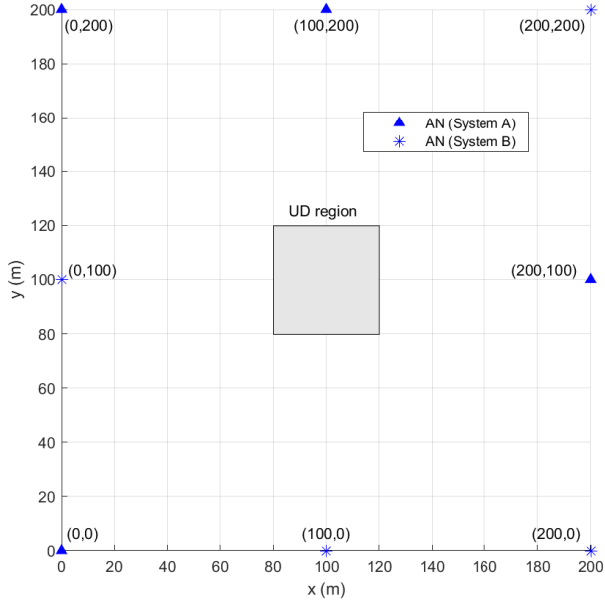}
	\caption{AN placement and UD position for 2D simulation scene.
	}
	\label{fig:Simsetting}
\end{figure}

We first create a 2D simulation scene with 4 ANs from system $A$ and 4 ANs from system $B$. As shown in Fig. \ref{fig:Simsetting}, the ANs are placed on a plain at the sides and corners of a square area with a side length of 200 m. All the positions of ANs are known without error. UD is placed randomly in a square region with a side length of 40 m. To ensure the far field assumption for the proposed method, the UD area is placed in the middle of the area. We set the $\sigma$ of the pseudorange measurement noise varying from 0.1 m to 10 m with a step of 0.9 m. Thus, there are 12 steps in total. At every step, we conduct 1,500 Monte-Carlo simulations with uniformly distributed random positions of UD inside the gray region.

\begin{figure}
	\centering
	\includegraphics[width=0.99\linewidth]{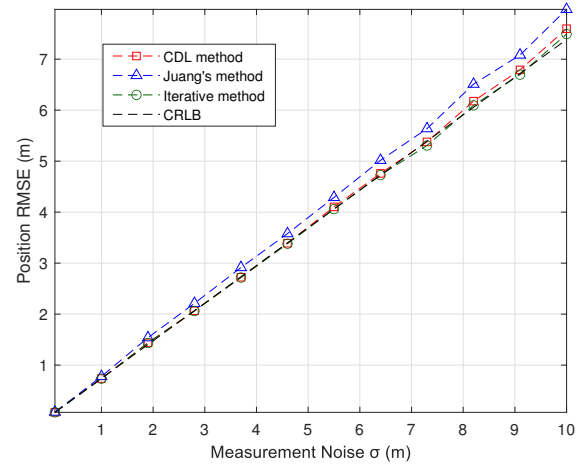}
	\caption{Position RMSE vs. measurement noise in 2D simulation scene. The new CDL method reaches CRLB with small noise. Positioning accuracy of the new CDL method is close to the iterative method and is better than that of the Juang's method.
	}
	\label{fig:2Derrorvsnoise}
\end{figure}

The position error result with varying measurement noise is depicted in Fig. \ref{fig:2Derrorvsnoise}. The errors from the Juang's method and iterative method are included for comparison. Theoretical position error lower bound from CRLB is computed based on (\ref{eq:0505CRLB2D}). It can be observed that, under the condition of small noise and far field, the positioning accuracy of the proposed method reaches CRLB. The position error of the proposed method is close to that of the iterative method and is smaller than that of the Juang's method throughout the measurement noise varying range. When the measurement noise increases, all three methods show degraded positioning accuracy. The proposed method performs slightly worse than the iterative method in terms of positioning accuracy, but still outperforms the Juang's method since its RMSE is closer to CRLB. This result validates the feasibility of the proposed method in dual-system case and verifies the theoretical error analysis in the previous section.

To evaluate computational complexity, we compare the running time of the new CDL method, the conventional iterative approach and the Juang's method. The Monte-Carlo simulation consists of 18,000 calls for each algorithm. The total running times for the Monte-Carlo simulation run of the new CDL method, the Juang's method and the iterative method are 3.38 s, 3.48 s and 6.49 s, respectively. The computation time of the new CDL method is the least among the three methods. Compared with the iterative method, the low complexity of the new CDL method mainly attributes to the non-iterative feature of the proposed method. Additional Monte-Carlo simulations give consistent results showing that the new CDL method has the least computational complexity. Thus, from Fig. \ref{fig:2Derrorvsnoise}, it can be seen that the CDL method can obtain similar positioning accuracy with much lower computational complexity compared with the conventional iterative method.

\subsection{3D Simulation}
\begin{figure}
	\centering
	\includegraphics[width=1.0\linewidth]{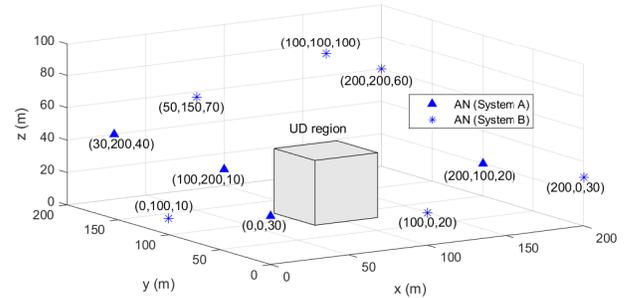} 
	\caption{AN placement and UD position for 3D simulation scene.
	}
	\label{fig:3DSimsetting}
\end{figure}

A 3D simulation scene is created to evaluate the positioning performance of the new CDL method in 3D case. There are 4 ANs from system $A$ and 6 ANs from system $B$. UD is placed randomly in a cubic region with a size of 40 m $\times$ 40 m $\times$ 40 m centered at (100, 100, 20) m. The locations of ANs and UD are shown in Fig. \ref{fig:3DSimsetting}. The range measurement noise $\sigma$ is varying from 0.1 m to 10 m with a step of 0.9 m in this simulation. 1,500 Monte-Carlo simulations with a random position of UD inside the UD region are conducted for each step.

\begin{figure}
	\centering
	\includegraphics[width=0.99\linewidth]{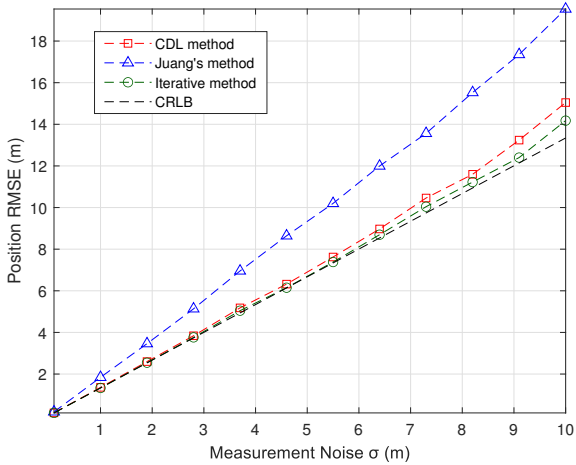} 
	\caption{Position RMSE vs. measurement noise in 3D simulation scene. The new CDL and the iterative method reach CRLB with small noise. The new CDL outperforms the Juang's method in terms of positioning accuracy.
	}
	\label{fig:3Derrorvsnoise}
\end{figure}

The position errors of the CDL method are illustrated in Fig. \ref{fig:3Derrorvsnoise}. It can be seen that, with small measurement noise, the localization error of the new CDL method reaches CRLB. The localization error of the CDL method is similar to that of the iterative method and is closer to CRLB than that of the Juang's method. This numerical localization result also matches the error analysis in the previous section.

The computation times for the new CDL method, the Juang's method and the iterative method are 3.51 s, 3.72 s and 7.95 s, respectively. The computation time of the new CDL method is the least among the three methods, identical with the result in the 2D simulation. This indicates a significant reduction in complexity of the new CDL algorithm compared with the iterative method. 

To summarize the 3D simulation, the numerical results also verify that the positioning accuracy of the new CDL method reaches CRLB under small noise and far field condition. With increasing measurement noise, its performance degrades but is still closer to CRLB than the Juang's method. Besides, its computational complexity is smaller than that of the iterative method.

\subsection{Real GPS+BDS Data Experiment}
\begin{figure}
	\centering
	\includegraphics[width=1\linewidth]{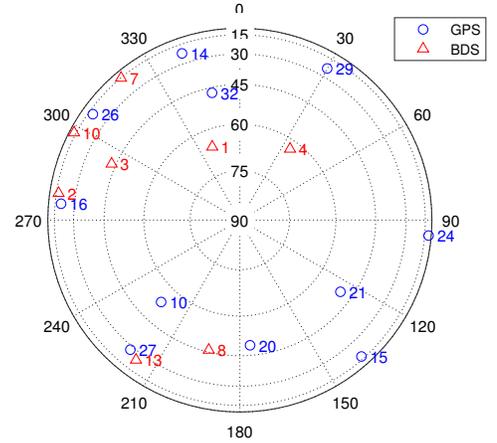} 
	\caption{Sky view of visible GPS and BDS satellites at one epoch of the real-world observation data.
	}
	\label{fig:skyplotGPSBDS}
\end{figure}

In order to evaluate the performance in real-world applications, we implement the CDL algorithm to process real GNSS pseudorange observation data. The iterative method is also realized as comparison. A 24-hour consecutive GPS and BDS real observation data set with a 30 s sampling interval from IGS Site TOW2, Cape Ferguson, Australia, is used. The observation period starts from 0:00, October 1, and ends at 0:00, October 2, 2018 (Universal Time Coordinated). These observation data are available on BKG Data Center website \cite{BKGcenter}. The navigation message data covering the same period  from Crustal Dynamics Data Information System (CDDIS) website \cite{CDDIS} are used to calculate the satellite positions. The sky view of the visible GPS and BDS satellites at one epoch of the data is depicted in Fig. \ref{fig:skyplotGPSBDS}.

\begin{figure*}[!t]
\centering
\subfloat[CDL method]{\includegraphics[width=0.5\linewidth]{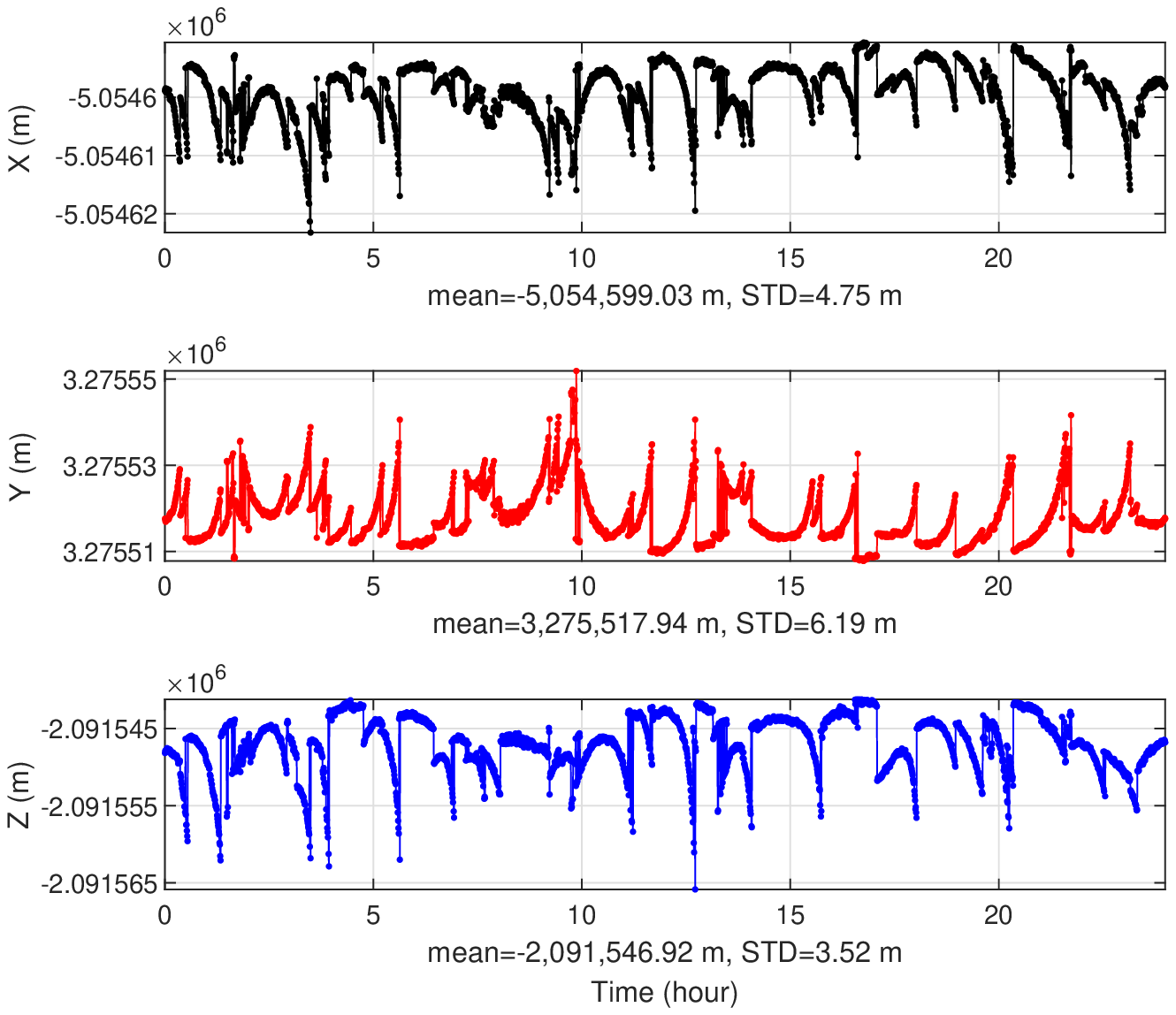}%
\label{fig_proposed}}
\hfil
\subfloat[Iterative method]{\includegraphics[width=0.5\linewidth]{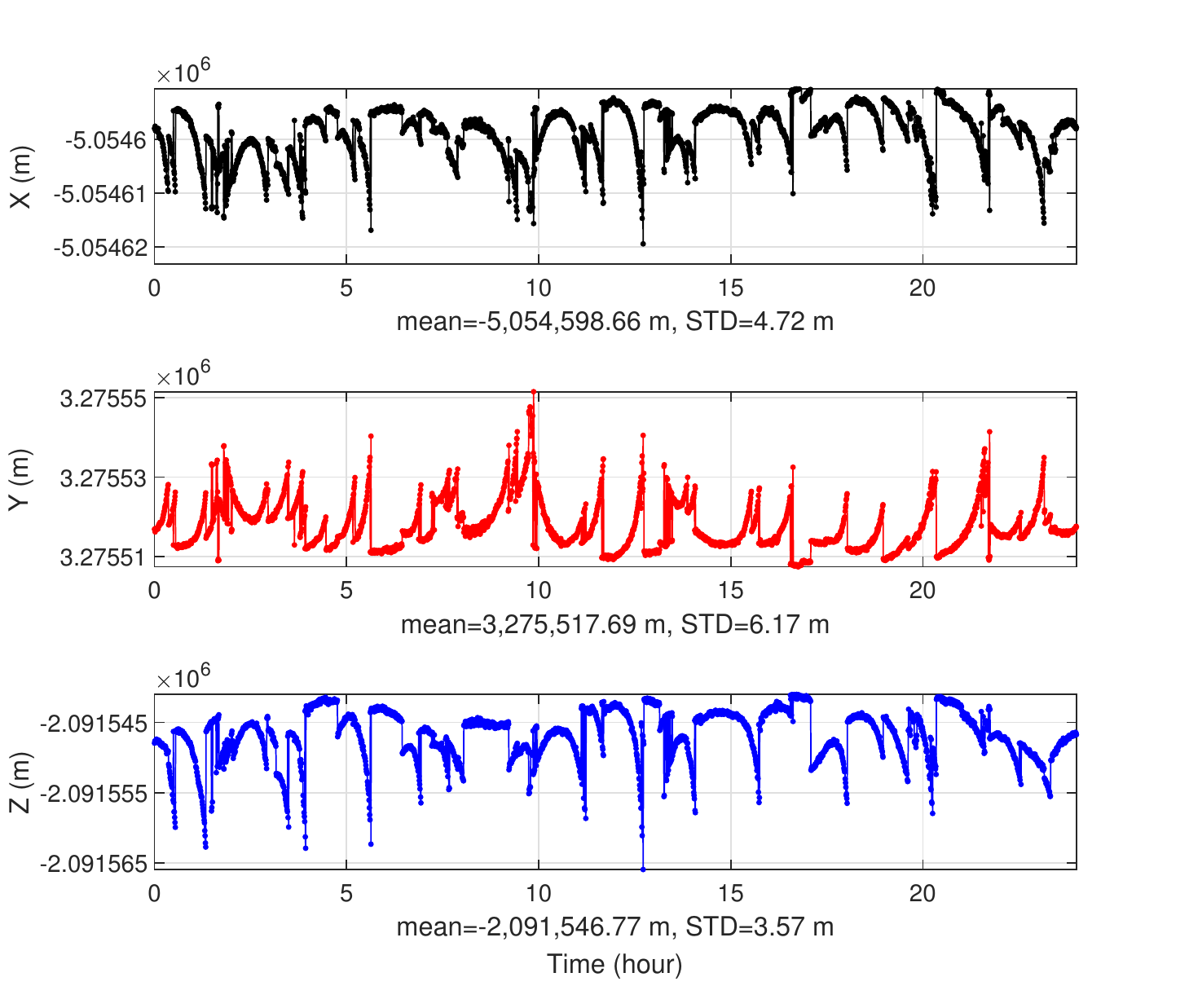}%
\label{fig_iter}}
\caption{Three-axis positioning results of the CDL and iterative method. Positioning result of each axis from the new CDL method is essentially the same as that of the iterative method}
\label{fig:realGPSBDS}
\end{figure*}

The 3D positioning results in the earth-centered, earth-fixed (ECEF) coordinate of both the CDL algorithm and the iterative method are shown in Fig. \ref{fig:realGPSBDS}.
It can be seen that 3-axis positioning results including both the mean coordinate and the standard deviation (STD) of both methods are almost identical. The localization result curves for all three axes of both methods have an identical epoch-by-epoch pattern, showing that the two methods have almost the same localization accuracy. The similarity of the localization performance between the CDL method and the conventional iterative approach is consistent with the simulated 2D and 3D results in the above sub-sections. This validates the feasibility and performance of the new CDL method in the real world.

The computational complexity is also evaluated. The real-world data set has 2,880 epochs in total. That means the new CDL algorithm and the iterative method are respectively called 2,880 times when processing the real data. The computation time cost of the new CDL method is 1.93 s compared with 3.64 s for the iterative method, about 40\% improvement. This shows a complexity reduction with comparable positioning accuracy of the CDL method in the real-world application compared with the conventional iterative method.

With the fast development of IoT, more and more new applications such as drone control, vehicle positioning \& navigation, and location-based services require higher accuracy and better availability. Dual localization systems, such as GPS and BDS, can be used to meet such requirements by adopting the new CDL method for these novel applications. Besides, low computation complexity of the new CDL method as shown in the experiment can benefit these applications on size and power-constrained electronics systems such as cell-phones, digital bracelets, and drone platforms.

\section{Conclusion}
In this paper, a new closed-form dual-system localization algorithm, namely CDL, for two non-synchronized pseudorange based systems is developed. In this method, the non-linear relationship between the user position and the pseudorange measurements is converted to a linear one by taking a square on the distances. Solving for the user position is then reduced to finding the roots to two intermediate distance variables in a closed form. After analytically solving a quadratic equation set to identify the intermediate variables, the user position is computed by applying a WLS method. Theoretical analysis on the localization error covariance of the new CDL method is conducted. We prove that the positioning accuracy reaches CRLB under small noise and far field condition. Compared with the iterative method, the new CDL method does not require initial guess and has lower complexity with similar positioning accuracy. The localization accuracy is better than the state-of-the-art closed-form dual-system method. Simulations in 2D and 3D scenes verify the theoretical analysis that the positioning accuracy of the new method reaches CRLB and is superior over the existing closed-form dual-system approach. It is also verified that the new CDL method has much lower computational cost than the iterative method with comparable positioning accuracy. Experiments using real GPS and BDS data are conducted. The results show that the CDL method is feasible in real-world GNSS applications and its computational complexity can be reduced by about 40\%.


\appendices
\section{Solution to Quadratic Equation Set}
\label{Appendix1}
By replacing the unknowns with $x$ and $y$, respectively, we rewrite the two quadratic equations of (\ref{eq:0319quadratic1}) and (\ref{eq:0320quadratic2}) as 
\begin{equation} \label{eq:A001quadratic1}
\begin{split}
a_1 x^2+b_1 x y +c_1 y^2 + d_1 x+e_1 y+f_1=0 \text{,}
\end{split}
\end{equation}
\begin{equation} \label{eq:A002quadratic2}
\begin{split}
a_2 x^2+b_2 xy +c_2 y^2 + d_2 x+e_2 y+f_2=0 \text{.}
\end{split}
\end{equation}

We first remove the $y^2$ term by multiplying $c_1$ and $c_2$ to (\ref{eq:A002quadratic2}) and (\ref{eq:A001quadratic1}), respectively, and then subtracting the resulting equations. After re-organizing, we obtain
\begin{equation} \label{eq:A004yequation}
(t_{1}x+t_2)y=t_3 x^2+t_4 x+t_5 \text{,}
\end{equation}
where
$$
t_1=b_1 c_2-b_2 c_1 \text{,} \; t_2=e_1 c_2 -e_2 c_1 \text{,}
$$
$$
t_3=-a_1 c_2+ a_2 c_1, \; t_4 = -d_1 c_2 + d_2 c_1 \text{,}
$$
$$
t_5=-f_1 c_2 + f_2 c_1 \text{.}
$$

Here are two cases. One is $t_{1}x+t_2=0 $ and the other is $t_{1}x+t_2 \neq 0 $.

Case 1: $t_{1}x+t_2=0 $

If $t_1=0$, then $t_2$ must equal to zero. In this sub-case, the problem reduces to solving the equation of 
\begin{equation} \label{eq:A005xequation}
t_3 x^2+t_4 x+t_5=0 \text{.}
\end{equation}

After substituting the root of $x$ from (\ref{eq:A005xequation}) into (\ref{eq:A002quadratic2}), the root of $y$ can be found.

If $t_1\neq 0$, then we need to test if $-t_2/t_1$ is the root of $x$ by substituting it into (\ref{eq:A005xequation}). If it satisfies (\ref{eq:A005xequation}), then the root of $y$ can be found by substituting $x$ into (\ref{eq:A002quadratic2}). Otherwise, there is no solution.

Case 2: $t_{1}x+t_2\neq 0 $

In this case, we have 
\begin{equation} \label{eq:A007commoncase}
y=(t_3 x^2+t_4 x+t_5)/(t_{1}x+t_2) \text{.}
\end{equation}

By substituting (\ref{eq:A007commoncase}) into (\ref{eq:A002quadratic2}), we come to a quartic equation of $x$ as
\begin{equation} \label{eq:A009quartic}
\alpha x^4+\beta x^3+ \gamma x^2 +\lambda x +\mu=0 \text{,}
\end{equation}
where
$$
\alpha=a_1 t_1^2 + b_1 t_1 t_3 + c_1 t_3^2 \text{,}
$$
$$
\beta=d_1 t_1^2 + 2 a_1 t_1 t_2 + b_1 t_1 t_4 + b_1 t_2 t_3 + 2 c_1 t_3 t_4 + e_1 t_1 t_3 \text{,}
$$
$$
\begin{array}{rr}
\gamma =c_1 (t_4^2 + 2 t_3 t_5) + a_1 t_2^2 + f_1 t_1^2 + b_1 t_1 t_5\\
+ b_1 t_2 t_4 + 2 d_1 t_1 t_2 + e_1 t_1 t_4 + e_1 t_2 t_3 \text{,}
\end{array} 
$$
$$
\lambda=d_1 t_2^2 + b_1 t_2 t_5 + 2 c_1 t_4 t_5 + e_1 t_1 t_5 + e_1 t_2 t_4 + 2 f_1 t_1 t_2 \text{,}
$$
$$
\mu=f_1 t_2^2 + e_1 t_2 t_5 + c_1 t_5^2 \text{.}
$$

The closed-form solution of the quartic equation can be found in mathematical literature such as \cite{shmakov2011universal,riccardi2009solution}. We simply write the solution as follows in this section without derivation so that interested readers are able to grasp the final result without diving into literature. There are at most four roots for this equation, either real or complex values. The general form of the roots is given by
\begin{equation} \label{eq:A010xroots}
\begin{split}
x(1), x(2)=-\frac{\beta}{4\alpha} - s \pm \frac{1}{2}\sqrt{-4s^2-2p+\frac{q_0}{s}} \text{,}\\
x(3), x(4)=-\frac{\beta}{4\alpha} + s \pm \frac{1}{2}\sqrt{-4s^2-2p-\frac{q_0}{s}} \text{.}
\end{split}
\end{equation}
with the variables expressed as follows,
$$
s=\frac{1}{2} \sqrt{-\frac{2}{3}p+\frac{1}{3\alpha} (q_1+\frac{\Delta_0}{q_1})}\text{,}
\;p=\frac{8 \alpha \gamma - 3 \beta^2}{8 \alpha^2} \text{,}
$$
$$
q_0=\frac{\beta^3 - 4 \alpha \beta \gamma + 8 \alpha^2 \lambda}{8 \alpha^3}\text{,} \;q_1=\sqrt[3]{\frac{\Delta_1+\sqrt{-27\Delta}}{2}} \text{,}
$$
$$
\Delta = -\frac{\Delta_1^2-4\Delta_0^3}{27}\text{,}\;\Delta_0=\gamma^2-3 \beta \lambda +12 \alpha \mu \text{,}
$$
$$
\Delta_1=2 \gamma^3 -9 \beta \gamma \lambda +27 \beta^2 \mu+27 \alpha \lambda^2 -72\alpha \gamma \mu \text{.}
$$

\section{Proof of Theorem 1}
\label{Appendix2}
	Let $\hat{\bm{C}}$ and $\hat{\boldsymbol{h}}$ be the noisy versions of $\bm{C}$ and $\boldsymbol{h}$ in (\ref{eq:0307Puvsrvector}), respectively.
The error vector $\boldsymbol\psi$ is then defined as
\begin{equation} \label{eq:0311errorvec}
\boldsymbol\psi \triangleq \hat{\bm{C}} \left[r_{A_1},  r_{B_1} \right] +\hat{\boldsymbol{h}}-\bm{G}\boldsymbol{p}_{u}.
\end{equation}

The weighting matrix $\bm{W}$ can be written in terms of the covariance of the error vector $\boldsymbol\psi$ as
\begin{equation} \label{eq:0311Wandcoverror}
\bm{W}=\mathbb{E}\left[\boldsymbol\psi\boldsymbol\psi^T\right]\text{.}
\end{equation}

The first row of $\hat{\bm{C}}$ and the first element of $\hat{\boldsymbol{h}}$ in (\ref{eq:0311errorvec}) are given by
\begin{equation} \label{eq:0313C_hatrow}
[\hat{\bm{C}}]_{1,:} = [r_{A_2A_1}+\epsilon_{A_2A_1}, 0] \text{,}
\end{equation}
and
\begin{equation} \label{eq:0314h_hatrow}
[\hat{\boldsymbol{h}}]_{1}=\frac{1}{2}\left(\left(r_{A_2A_1}+\epsilon_{A_2A_1}\right)^2+\Vert \boldsymbol{p}_{A_1}\Vert ^2 -\Vert \boldsymbol{p}_{A_2}\Vert ^2\right),
\end{equation}
respectively.

The first element of $\boldsymbol\psi$ is then given by
\begin{align} \label{eq:0316errorrow}
[\boldsymbol\psi]_{1}&=[\hat{\bm{C}}]_{1,:}\left[r_{A_1},  r_{B_1} \right]^T+[\hat{\boldsymbol{h}}]_{1}-[\bm{G}]_{1,:}\boldsymbol{p}_{u} \nonumber\\
& \;\;\;\; + \epsilon_{A_2A_1}(r_{A_1}+r_{A_2A_1})+\frac{1}{2}\epsilon_{A_2A_1}^2 \nonumber\\
&=\epsilon_{A_2A_1}r_{A_2}+\frac{1}{2}\epsilon_{A_2A_1}^2 \text{.}
\end{align}

Given the condition that UD is far from ANs and the measurement noise is small, the second squared error term in (\ref{eq:0316errorrow}) can be ignored, i.e.,
\begin{equation} \label{eq:0317errorrowappro}
\begin{split}
[\boldsymbol\psi]_{1}=\epsilon_{A_2A_1}r_{A_2}.
\end{split}
\end{equation}

The vector form of $\boldsymbol\psi$ is then written as
\begin{equation} \label{eq:0318vectorpsi}
\boldsymbol\psi= \left[
\begin{matrix}
r_{A_2}(\epsilon_{A_2}-\epsilon_{A_1})\\
\vdots\\
r_{A_M}(\epsilon_{A_M}-\epsilon_{A_1})\\
r_{B_2}(\epsilon_{B_2}-\epsilon_{B_1})\\
\vdots\\
r_{B_N}(\epsilon_{B_N}-\epsilon_{B_1})
\end{matrix}
\right]
=\bm{D}\bm{\zeta}\text{,}
\end{equation}
where $
\bm{D}= \mathrm{diag} \left(r_{A_2},\cdots,r_{A_M},r_{B_2},\cdots,r_{B_N}\right)
$,
$$
\bm{\zeta}=\left[
\begin{matrix}
\epsilon_{A_2}-\epsilon_{A_1}\\
\vdots\\
\epsilon_{A_M}-\epsilon_{A_1}\\
\epsilon_{B_2}-\epsilon_{B_1}\\
\vdots\\
\epsilon_{B_N}-\epsilon_{B_1}
\end{matrix}\right] \text{.} 
$$

Then, the covariance of $\boldsymbol\psi$ is given by

\begin{equation} \label{eq:0320coverror0}
\mathbb{E}\left[\boldsymbol\psi\boldsymbol\psi^T\right]=\bm{D}\mathbb{E}\left[
\bm{\zeta}\bm{\zeta}^T\right]
\bm{D} \text{.}
\end{equation}

We note that the pseudorange measurement noises for all ANs are i.i.d. and follow a Gaussian distribution. Therefore, 
$$
\mathbb{E}\left[\epsilon_{A_i}^2\right]=\sigma_{A_i}^2 \text{,} \; \mathbb{E}\left[\epsilon_{B_j}^2\right]=\sigma_{B_j}^2 \text{,}
$$
and the covariance terms between different pseudoranges and different systems are zero, i.e.,
$$
\mathbb{E}\left[\epsilon_{A_i}\epsilon_{A_m}\right]=\mathbb{E}\left[\epsilon_{B_j}\epsilon_{B_n}\right]=\mathbb{E}\left[\epsilon_{A_i}\epsilon_{B_j}\right]=0 \text{,} \; i\neq m, j\neq n \text{.}
$$

We denote the expectation term on the right side of (\ref{eq:0320coverror0}) as $\bm{Q}$, which then has the form of
$$
\bm{Q}=\begin{bmatrix}
\bm{Q}_A
& \bm{O}_{(M-1) \times (N-1)} \\
\bm{O}_{(N-1) \times (M-1)} &
\bm{Q}_B
\end{bmatrix}
\text{,}
$$
where,
$$
\bm{Q}_A =
\left[
\begin{matrix}
\sigma_{A_1}^2+ \sigma_{A_2}^2 &\sigma_{A_1}^2 & \cdots &\sigma_{A_1}^2  \\
\sigma_{A_1}^2 &  \sigma_{A_1}^2+ \sigma_{A_3}^2 &\cdots &\sigma_{A_1}^2\\
\vdots & \vdots & \ddots & \vdots \\
\sigma_{A_1}^2 & \sigma_{A_1}^2 & \cdots &\sigma_{A_1}^2+ \sigma_{A_M}^2
\end{matrix}
\right]\text{,}
$$
and
$$
\bm{Q}_B =
\left[
\begin{matrix}
\sigma_{B_1}^2+ \sigma_{B_2}^2 &\sigma_{B_1}^2 & \cdots &\sigma_{B_1}^2  \\
\sigma_{B_1}^2 & \sigma_{B_1}^2+ \sigma_{B_3}^2 & \cdots & \sigma_{B_1}^2 \\
\vdots & \vdots & \ddots & \vdots \\
\sigma_{B_1}^2 & \sigma_{B_1}^2 &\cdots &\sigma_{B_1}^2+ \sigma_{B_N}^2
\end{matrix}
\right] \text{.}
$$	

As a result, the covariance matrix of $\boldsymbol\psi$ is written as
\begin{equation} \label{eq:0319coverror}
\mathbb{E}\left[\boldsymbol\psi\boldsymbol\psi^T\right]=\bm{D}\bm{Q}\bm{D} \text{.}
\end{equation}

Finally, based on (\ref{eq:0311Wandcoverror}) and (\ref{eq:0319coverror}), we have obtained the expression of the weighting matrix $\bm{W}$, which is identical with (\ref{eq:0316Weightmatrix}). Thus, we have finished the proof of Theorem 1.

\section{Derivation of Simplified Solution Selection Form}\label{Appendix3}
Based on (\ref{eq:0316Weightmatrix}), matrix $\bm{Q}$ is rewritten as
\begin{align} \label{eq:A003Qmat}
\bm{Q} =
&\mathrm{diag}\left(\sigma_{A_2}^2,\cdots,\sigma_{A_M}^2,\sigma_{B_2}^2,\cdots,\sigma_{B_N}^2\right) \nonumber \\
&+\left[
\begin{matrix}
\sigma_{A_1}^2\bm{J}_{(M-1)\times (M-1)} & \bm{O}_{(M-1)\times (N-1)}  \\
\bm{O}_{(N-1)\times (M-1)} &  \sigma_{B_1}^2\bm{J}_{(N-1)\times (N-1)}
\end{matrix}
\right] \text{,}
\end{align}
where $\bm{J}$ is a matrix with all entries being one.

According to \cite{sathyan2010analysis}, the inverse of $\bm{Q}$ is written as
\begin{align} \label{eq:A004invQmat}
\bm{Q}^{-1} =
&\mathrm{diag}\left(\frac{1}{\sigma_{A_2}^2},\cdots,\frac{1}{\sigma_{A_M}^2},\frac{1}{\sigma_{B_2}^2},\cdots,\frac{1}{\sigma_{B_N}^2}\right) \nonumber \\
&-\left[
\begin{matrix}
\bm{X}_{A} & \bm{O}_{(M-1)\times (N-1)}  \\
\bm{O}_{(N-1)\times (M-1)} &  \bm{X}_{B}
\end{matrix}
\right] \text{,}
\end{align}
where
$$
\bm{X}_{A}=
\frac{1}{\sum_{i=1}^{M}\frac{1}{\sigma_{A_i}^2}}
\left[
\begin{matrix}
\frac{1}{\sigma_{A_2}^4} & \frac{1}{\sigma_{A_2}^2\sigma_{A_3}^2} &\cdots &\frac{1}{\sigma_{A_2}^2\sigma_{A_M}^2} \\
\frac{1}{\sigma_{A_3}^2\sigma_{A_2}^2} &  \frac{1}{\sigma_{A_3}^4} &\cdots &\frac{1}{\sigma_{A_3}^2\sigma_{A_M}^2} \\
\vdots&\vdots&\ddots&\vdots  \\
\frac{1}{\sigma_{A_M}^2\sigma_{A_2}^2}&\frac{1}{\sigma_{A_M}^2\sigma_{A_3}^2} &\cdots &\frac{1}{\sigma_{A_M}^4}
\end{matrix}
\right] \text{,}
$$
and
$$
\bm{X}_{B}=
\frac{1}{\sum_{i=1}^{N}\frac{1}{\sigma_{B_i}^2}}
\left[
\begin{matrix}
\frac{1}{\sigma_{B_2}^4} & \frac{1}{\sigma_{B_2}^2\sigma_{B_3}^2} &\cdots &\frac{1}{\sigma_{B_2}^2\sigma_{B_N}^2} \\
\frac{1}{\sigma_{B_3}^2\sigma_{B_2}^2} &  \frac{1}{\sigma_{B_3}^4} &\cdots &\frac{1}{\sigma_{B_3}^2\sigma_{B_N}^2} \\
\vdots&\vdots&\ddots&\vdots  \\
\frac{1}{\sigma_{B_N}^2\sigma_{B_2}^2}&\frac{1}{\sigma_{B_N}^2\sigma_{B_3}^2} &\cdots &\frac{1}{\sigma_{B_N}^4}
\end{matrix}
\right] \text{.}
$$

The second matrix term of (\ref{eq:A004invQmat}) can be ignored if the number of ANs, i.e., $M$ and $N$, are large. Furthermore, if all the measurement noise variances are identical and denoted by $\sigma^2$, then $\bm{Q}^{-1} \approx \frac{1}{\sigma^2}\bm{I}_{(M+N-2)\times (M+N-2)}$, where $\bm{I}$ is an identity matrix. Therefore, (\ref{eq:0322residual}) can reduce to the form of $\min_{\tilde{\boldsymbol{p}}_u} \boldsymbol{d}_{\rho}^T\boldsymbol{d}_{\rho}$.

\section{Proof of Equivalence for Position Related CRLB using TOA and TDOA from Dual Systems}\label{Appendix4}
In the dual-system positioning case, when using TDOA measurements, the FIM is written as
\begin{equation} \label{eq:A0101FIMdualTDOA}
\bm{F}_{TDOA}=\bm{H}^T\bm{Q}^{-1}\bm{H} \text{.}
\end{equation}
where $\bm{H}$ and $\bm{Q}$ have the same definitions as in (\ref{eq:0417delPuvector}) and (\ref{eq:0316Weightmatrix}), respectively.

If we divide $\bm{H}$ column-wisely into two blocks relating to systems $A$ and $B$, respectively, then we have
\begin{equation} \label{eq:A0102HintoBlocks}
\bm{H}=\left[\bm{H}_A^T,\bm{H}_B^T\right]^T \text{,}
\end{equation}
where
$$
\bm{H}_A=
\begin{bmatrix}
\boldsymbol{l}_{A_1}^T-\boldsymbol{l}_{A_2}^T \\
\vdots \\
\boldsymbol{l}_{A_1}^T-\boldsymbol{l}_{A_M}^T \\
\end{bmatrix}
\text{,}\;
\bm{H}_B=
\begin{bmatrix}
\boldsymbol{l}_{B_1}^T-\boldsymbol{l}_{B_2}^T \\
\vdots \\
\boldsymbol{l}_{B_1}^T-\boldsymbol{l}_{B_N}^T \\
\end{bmatrix}\text{.}
$$

We note that $\bm{Q}$ is divided into blocks in  (\ref{eq:0316Weightmatrix}). Then, the FIM is rewritten as
\begin{equation} \label{eq:A0103FIMblock}
\bm{F}_{TDOA}=\bm{H}_A^T\bm{Q}_A^{-1}\bm{H}_A+\bm{H}_B^T\bm{Q}_B^{-1}\bm{H}_B \text{.}
\end{equation}

When using TOA or pseudorange measurements for the dual-system positioning case, the FIM denoted by $\bm{F}_{TOA}$ is written as
\begin{equation}
\begin{split}
\bm{F}_{TOA}=\bm{H}_{TOA}^T\bm{Q}_{TOA}^{-1}\bm{H}_{TOA} 
=\begin{bmatrix}
\bm{F}_{11} & \bm{F}_{12} \\
\bm{F}_{12}^T & \bm{F}_{22}
\end{bmatrix} \text{,}
\end{split}
\end{equation}
where
$$
\bm{H}_{TOA}=
\begin{bmatrix}
\bm{H}_{A_{TOA}} & -\bm{1}_M &\bm{0}_M \\
\bm{H}_{B_{TOA}} & \bm{0}_N  &-\bm{1}_N 
\end{bmatrix}\text{,}
$$
$$
\bm{Q}_{TOA}=
\begin{bmatrix}
\bm{Q}_{A_{TOA}} & \bm{O}_{M \times N} \\
\bm{O}_{N \times M} & \bm{Q}_{B_{TOA}}
\end{bmatrix} \text{,}
$$
$$
\bm{H}_{A_{TOA}}=
\begin{bmatrix}
\boldsymbol{l}_{A_1}^T & -1 & 0 \\
\vdots &\vdots&\vdots\\
\boldsymbol{l}_{A_M}^T & -1 & 0 \\
\end{bmatrix}
\text{,}\;
\bm{H}_{B_{TOA}}=
\begin{bmatrix}
\boldsymbol{l}_{B_1}^T & 0 & -1 \\
\vdots &\vdots&\vdots \\
\boldsymbol{l}_{B_N}^T & 0 & -1 \\
\end{bmatrix} \text{,}
$$
$$
\bm{Q}_{A_{TOA}}= \mathrm{diag} (\sigma_{A_1}^2,\cdots, \sigma_{A_M}^2 )\text{,}
$$
$$
\bm{Q}_{B_{TOA}}= \mathrm{diag} (\sigma_{B_1}^2,\cdots, \sigma_{B_N}^2 )\text{,}
$$
$$
\bm{F}_{11}=
\bm{H}_{A_{TOA}}^T \bm{Q}_{A_{TOA}}^{-1} \bm{H}_{A_{TOA}} + \bm{H}_{B_{TOA}}^T \bm{Q}_{B_{TOA}}^{-1}\bm{H}_{B_{TOA}} \text{,}
$$
$$
\bm{F}_{12}=
\left[-\bm{H}_{A_{TOA}}^T \bm{Q}_{A_{TOA}}^{-1}\bm{1}_M ,-\bm{H}_{B_{TOA}}^T \bm{Q}_{B_{TOA}}^{-1}\bm{1}_N\right] \text{,}
$$
$$
\bm{F}_{22}=
\mathrm{diag}\left(\mathrm{tr}\left(\bm{Q}_{A_{TOA}}^{-1}\right),\mathrm{tr}\left(\bm{Q}_{B_{TOA}}^{-1}\right)\right) \text{.}
$$

The upper-left square sub-matrix (either $2\times2$ in 2D cases or $3\times 3$ in 3D cases) in the inverse of the TOA FIM ($\bm{F}_{TOA}^{-1}$) contains the CRLB relating to the position errors. We denote it by $\bm{J}_{pos}$. According to the inverse of a partitioned matrix \cite{horn2012matrix}, we come to
\begin{align} \label{eq:A0105Jpos}
&\bm{J}_{pos}^{-1}\nonumber\\
&=\bm{F}_{11}-\bm{F}_{12}\bm{F}_{22}^{-1}\bm{F}_{12}^T \nonumber\\
&=\bm{H}_{A_{TOA}}^T\bm{Q}_{A_{TOA}}^{-1}\bm{H}_{A_{TOA}}\nonumber\\
&\;\;\;\;-\bm{H}_{A_{TOA}}^T\bm{Q}_{A_{TOA}}^{-1}\bm{1}_M \mathrm{tr}(\bm{Q}_{A_{TOA}}^{-1})^{-1}\bm{1}_M^T\bm{Q}_{A_{TOA}}^{-1}\bm{H}_{A_{TOA}}\nonumber\\
&\;\;\;\;+\bm{H}_{B_{TOA}}^T\bm{Q}_{B_{TOA}}^{-1}\bm{H}_{B_{TOA}}\nonumber\\
&\;\;\;\;-\bm{H}_{B_{TOA}}^T\bm{Q}_{B_{TOA}}^{-1}\bm{1}_N \mathrm{tr}(\bm{Q}_{B_{TOA}}^{-1})^{-1}\bm{1}_N^T\bm{Q}_{B_{TOA}}^{-1}\bm{H}_{B_{TOA}} \text{.}
\end{align}

The problem then boils down to the proof of equivalence of $\bm{F}_{TDOA}$ and $\bm{J}_{pos}^{-1}$. By observing (\ref{eq:A0105Jpos}) and (\ref{eq:A0103FIMblock}), we notice that these two matrices are both the sum of system $A$ related terms and system $B$ related terms. If the terms of system $A$ (and $B$) in $\bm{F}_{TDOA}$ is equal to the $A$ (and $B$) related terms in $\bm{J}_{pos}^{-1}$, then the proof will be done. In other words, we need to prove that in the single-system case, the positioning CRLB using TOA measurements is identical with the one using TDOA measurements. This proof is presented in \cite{sathyan2010analysis,urruela2006average},
and interested readers are referred to their mathematical derivations.


%


\ifCLASSOPTIONcaptionsoff
    \newpage
\fi



\bibliographystyle{IEEEtran}
\bibliography{IEEEabrv,paper}
%


%








\begin{IEEEbiography}
[{\includegraphics[width=1in,height=1.25in,clip,keepaspectratio]{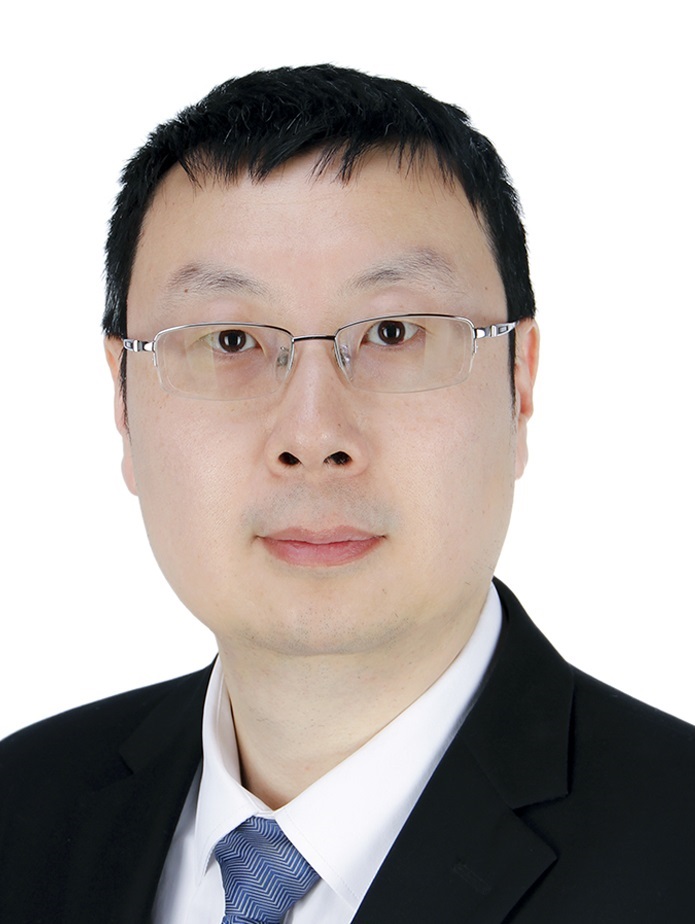}}]{Sihao Zhao}
received B.S. and Ph.D. degrees in Electronic Engineering from Tsinghua University, in 2005 and 2011, respectively.

From 2011 to 2013, he was an Electronics Systems Engineer with China Academy of Space Technology. From 2013 to 2019, he was a Postdoctoral Researcher and then an Assistant Professor with the Department of Electronic Engineering, Tsinghua University. Since 2020, he has been a Research Associate with the Communication and Signal Processing Applications Laboratory (CASPAL), Ryerson University. His research interests include localization algorithms, high-precision positioning techniques, and indoor navigation system development.
\end{IEEEbiography}
\vfill

\begin{IEEEbiography}
[{\includegraphics[width=1in,height=1.25in,clip,keepaspectratio]{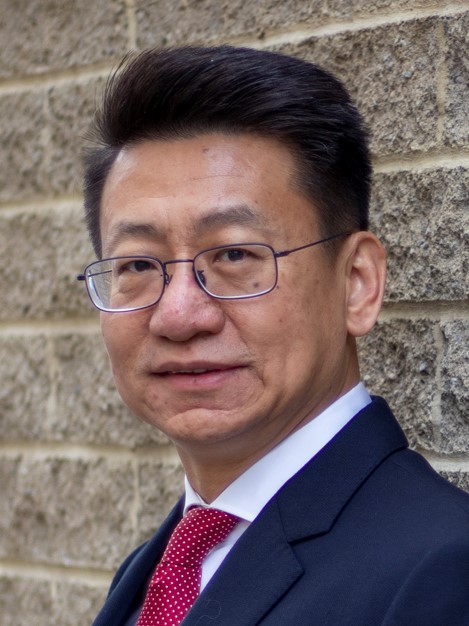}}]{Xiao-Ping Zhang}
received B.S. and Ph.D. degrees from Tsinghua University, in 1992 and 1996, respectively, both in Electronic Engineering. He holds an MBA in Finance, Economics and Entrepreneurship with Honors from the University of Chicago Booth School of Business, Chicago, IL. 

Since Fall 2000, he has been with the Department of Electrical and Computer Engineering, Ryerson University, Toronto, ON, Canada, where he is currently a Professor and the Director of the Communication and Signal Processing Applications Laboratory. He has served as the Program Director of Graduate Studies. He is cross-appointed to the Finance Department at the Ted Rogers School of Management, Ryerson University. He was a Visiting Scientist with the Research Laboratory of Electronics, Massachusetts Institute of Technology, Cambridge, MA, USA, in 2015 and 2017. He is a frequent consultant for biotech companies and investment firms. He is the Co-Founder and CEO for EidoSearch, an Ontario-based company offering a content-based search and analysis engine for financial big data. His research interests include sensor networks and IoT, machine learning, statistical signal processing, image and multimedia content analysis, and applications in big data, finance, and marketing. 

Dr. Zhang is a Fellow of Canadian Academy of Engineering, a registered Professional Engineer in Ontario, Canada, and a member of Beta Gamma Sigma Honor Society. He is the general Co-Chair for the IEEE International Conference on Acoustics, Speech, and Signal Processing, 2021. He is the general co-chair for 2017 GlobalSIP Symposium on Signal and Information Processing for Finance and Business, and the general co-chair for 2019 GlobalSIP Symposium on Signal, Information Processing and AI for Finance and Business. He is an elected Member of the ICME steering committee. He is the General Chair for the IEEE International Workshop on Multimedia Signal Processing, 2015. He is the Publicity Chair for the International Conference on Multimedia and Expo 2006, and the Program Chair for International Conference on Intelligent Computing in 2005 and 2010. He served as a Guest Editor for Multimedia Tools and Applications and the International Journal of Semantic Computing. He was a tutorial speaker at the 2011 ACM International Conference on Multimedia, the 2013 IEEE International Symposium on Circuits and Systems, the 2013 IEEE International Conference on Image Processing, the 2014 IEEE International Conference on Acoustics, Speech, and Signal Processing, the 2017 International Joint Conference on Neural Networks and the 2019 IEEE International Symposium on Circuits and Systems. He is a Senior Area Editor for the IEEE TRANSACTIONS ON SIGNAL PROCESSING and the IEEE TRANSACTIONS ON IMAGE PROCESSING. He was an Associate Editor for the IEEE TRANSACTIONS ON IMAGE PROCESSING, the IEEE TRANSACTIONS ON MULTIMEDIA, the IEEE TRANSACTIONS ON CIRCUITS AND SYSTEMS FOR VIDEO TECHNOLOGY, the IEEE TRANSACTIONS ON SIGNAL PROCESSING, and the IEEE SIGNAL PROCESSING LETTERS. He received 2020 Sarwan Sahota Ryerson Distinguished Scholar Award, the Ryerson University highest honor for scholarly, research and creative achievements. He is awarded as IEEE Distinguished Lecturer for the term from January 2020 to December 2021 by IEEE Signal Processing Society. 
\end{IEEEbiography}
\vspace{2cm}

\begin{IEEEbiography}[{\includegraphics[width=1in,height=1.25in,clip,keepaspectratio]{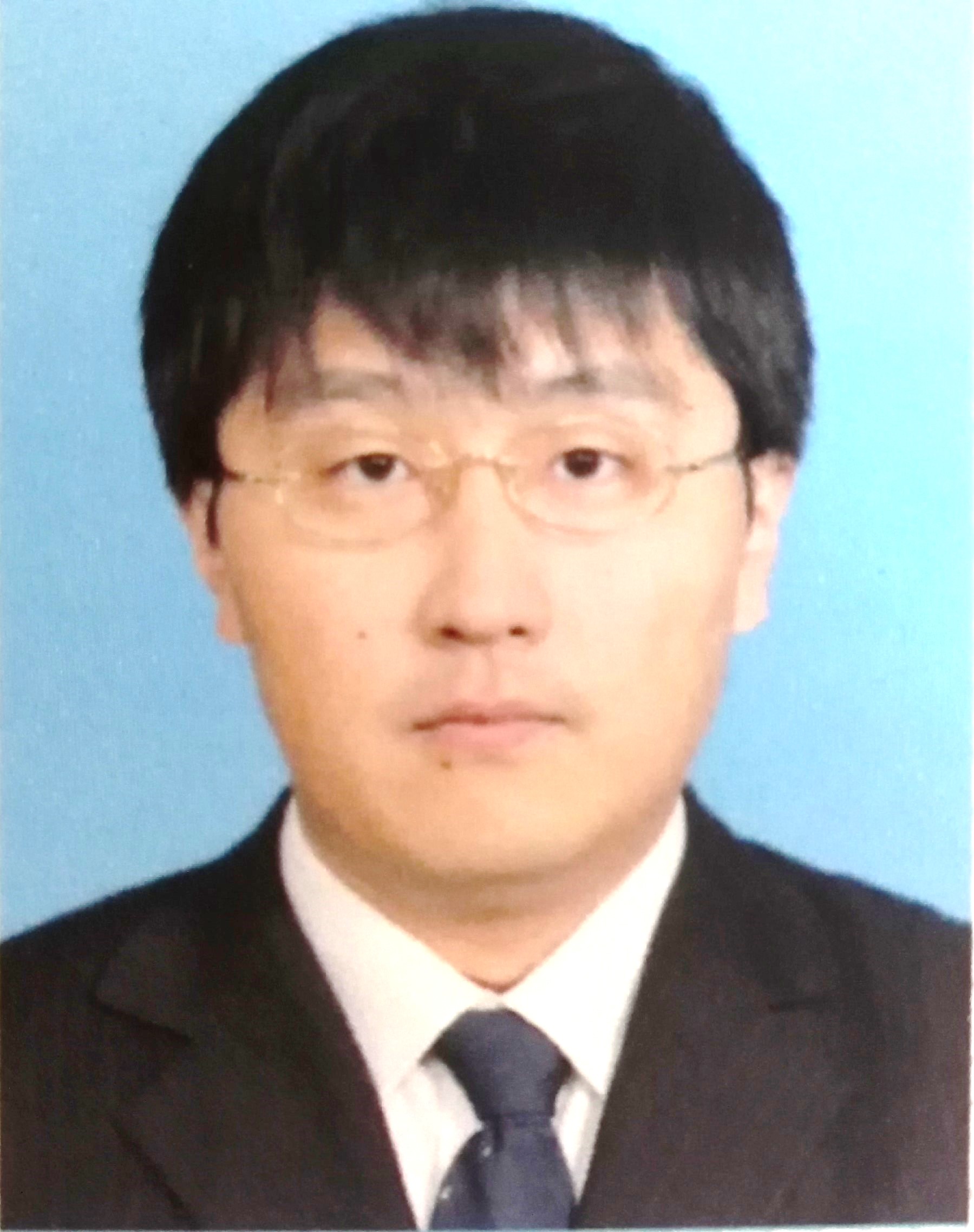}}]{Xiaowei Cui}
received B.S. and Ph.D. degrees in Electronic Engineering from Tsinghua University, in 2000 and 2005, respectively.

Since 2005, he has been with the Department of Electronic Engineering, Tsinghua University, Beijing, China, where he is an Associate Professor at present. He is a member of the Expert Group of China BeiDou Navigation Satellite System. His research interests include robust GNSS signal processing, multipath mitigation techniques and high-precision positioning.
\end{IEEEbiography}
\vspace{2cm}

\begin{IEEEbiography}[{\includegraphics[width=1in,height=1.25in,clip,keepaspectratio]{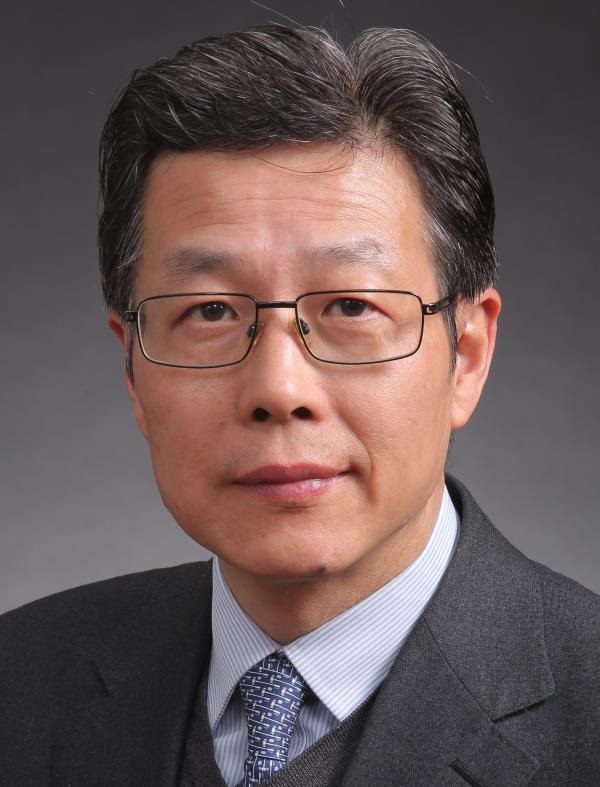}}]{Mingquan Lu}
received M.E. and Ph.D. degrees in Electronic Engineering from University of Electronic Science and Technology, Chengdu, China.

He is a Professor with the Department of Electronic Engineering, Tsinghua University, Beijing, China. He directs the Positioning, Navigation and Timing (PNT) Research Center, which develops GNSS and other PNT technologies. His current research interests include GNSS system modeling and simulation, signal design and processing, and receiver development. He is also a Research Fellow with Beijing National Research Center for Information Science and Technology.
\end{IEEEbiography}
\vspace{13cm}

\end{document}